\documentclass[preprint]{aastex}

\usepackage{natbib}
\usepackage[section]{placeins}

\shorttitle{Dynamics of subsurface active regions}
\shortauthors{Jouve, Brun \& Aulanier}

\begin{document}

\bibliographystyle{plainnat}

\title{GLOBAL DYNAMICS OF SUBSURFACE SOLAR ACTIVE REGIONS}

\author{L. Jouve\altaffilmark{1}}
\affil{UPS-OMP, Institut de Recherche en Astrophysique et Plan\'etologie, Universit\'e de Toulouse CNRS, 14 Avenue Edouard Belin, F-31400 Toulouse, France}

\email{ljouve@irap.omp.eu}

\and
\author{A. S. Brun\altaffilmark{2}}
\affil{Laboratoire AIM, CEA/DSM-CNRS-Universit\'e 
Paris Diderot, IRFU/SAp, F-91191 Gif sur Yvette, France}

\and 
\author{G. Aulanier}
\affil{LESIA, Observatoire de Paris, CNRS, UPMC, Universit\'e Paris-Diderot, 5 Place Jules Janssen, F-92190 Meudon Cedex, France}

\altaffiltext{1}{Associated to Laboratoire AIM, CEA/DSM-CNRS-Universit\'e
Paris Diderot, IRFU/SAp, F-91191 Gif sur Yvette, France}

\altaffiltext{2}{Associated to LESIA, Observatoire de Paris, CNRS, UPMC, Universit\'e Paris-Diderot, 5 Place Jules Janssen, F-92190 Meudon Cedex, France}

\begin{abstract}

We present three-dimensional numerical simulations of a magnetic loop evolving in either a convectively stable or unstable rotating shell. The magnetic loop is introduced in the shell in such a way that it is buoyant only in a certain portion in longitude, thus creating an $\Omega$-loop. Due to the action of magnetic buoyancy, the loop rises and develops asymmetries between its leading and following legs, creating emerging bipolar regions whose characteristics are similar to the ones of observed spots at the solar surface. In particular, we self-consistently reproduce the creation of tongues around the spot polarities, which can be strongly affected by convection. We moreover emphasize the presence of ring-shaped magnetic structures around our simulated emerging regions, which we call ``magnetic necklace'' and which were seen in a number of observations without being reported as of today. We show that those necklaces are markers of vorticity generation at the periphery and below the rising magnetic loop. We also find that the asymmetry between the two legs of the loop is crucially dependent on the initial magnetic field strength. The tilt angle of the emerging regions is also studied in the stable and unstable cases and seems to be affected both by the convective motions and the presence of a differential rotation in the convective cases.

\end{abstract}


\section{INTRODUCTION}

Our turbulent active Sun possesses a rather regular magnetic cycle that has been observed directly through the photospheric emergence of sunspots for about four centuries now. This magnetic flux emergence, which occurs daily at the Sun's surface, appears not only at large scales (sunspots and active regions) but also at scales as small as the intra-network field and there is indication that even smaller scale flux emerges at the solar surface, which remains unresolved with the present magnetographs \citep{Sanchez98, vanDriel02}. However, since a huge amount of flux is brought toward the photosphere by solar active regions and because of the well-established robust properties of such large-scale emergence, we focus our studies on the understanding of the dynamical evolution of large and strong magnetic structures rising from the solar interior to the surface. The appearance of active regions can be summarized in a few steps: small magnetic bipoles first emerge all over the future active region and then merge to form sunspots at the outer edges of the emerging flux region (EFR). The growing phase of a typical EFR lasts about 3-5 days. The flux in an active region is usually larger than $10^{21} \rm Mx$ in each polarity. The orientation of the bipole may be arbitrary at first, but generally in 1-3 days it becomes slightly tilted ($3^o$-$10^o$) with respect to the east-west direction, with the leading polarity closer to the equator, thus conforming to Joy's law \citep{Hale19}. The tilt angles of individual active regions show a large scatter about the mean but the measurements indicate a tendency for the mean value to linearly increase with the latitude of emergence \citep{Wang89,Wang91, DSilva93}. Moreover, various asymmetries have been detected in EFRs in particular in the morphology of the leading and trailing spots. Indeed, the leading polarity generally appears more intense and concentrated than the trailing one which tends to be more fragmented. Another interesting feature of EFRs is their inherent twist. Indeed, careful multi-wavelength analysis of flux emergence in active regions \citep[e.g.][]{Leka96} indicates that they appear with a certain amount of twist (i.e., a non-zero current along their axis). This twist is then thought to be responsible for the characteristic pattern known as tongues observed in newly emerging regions \citep{Lopez00} and identified as a major ingredient in mass ejections or flares at the solar photosphere \citep[e.g.,][]{Fan04b,Torok05, Aulanier10}. Other structures such as an annular shape of the magnetic field around active regions have been observed through various instruments \citep[e.g.,][]{Liu06} but never really reported as identified structures. However, as we will see, they can serve as strong constraints on models and simulations and the relevant physical processes responsible for their existence will be pointed out.

With all those observational constraints in mind, theoreticians and modelers have tried to analyze in detail the dynamical evolution of strong magnetic structures from the base of the convection zone (CZ) where they are created to the surface where they appear as sunspots. A possible scenario for such an evolution would be that strong toroidal fields are created at the base of the CZ by the differential rotation shearing a preexisting poloidal field in the tachocline (see landmark papers of \citealt{Moffatt78, Parker93}, recent simulations of \citealt{Browning06} and a review on the solar dynamo theory by \citealt*{Ossendrijver03} for example). After a certain period of storage in the subadiabatic layer below the CZ \citep{Moreno92, Schussler94}, this toroidal field becomes subject to magnetic buoyancy instabilities \citep{Parker55, Acheson79, Cattaneo88, Matthews95} and organizes itself in buoyant arched magnetic flux tubes (or $\Omega$-loops, \citealt{Zwann87}). The rise of those loops under the influence of Coriolis forces, turbulent convection and large-scale flows (rotation and meridional circulation) would then produce the wide variety of large-scale flux emergence observed at the solar surface, with the robust properties summarized above. Several numerical experiments of this last step have been performed using various levels of approximation, from the classical ``thin flux tube'' models \citep{Spruit81, Choudhuri87} to the more recent two-dimensional and three-dimensional calculations (see \citealt{Fan04a} for a detailed review on this subject). In particular, \citet{Emonet98} showed the necessity to start with a twisted magnetic field to counteract the vorticity generation within the flux tube and its subsequent breakup. \citet{Fan03} demonstrated the significant effect of convective motions on the deformation of the magnetic loop when the initial field was comparable to the equipartition field strength (corresponding to a magnetic energy equals to the kinetic energy of the strongest downflows). More recently, \citet{Jouve09} considered the effects of spherical geometry, convective motions, and self-consistently developed mean flows (differential rotation and meridional flow) on a rising flux tube. It was found that an initial twist of the field lines of about 14 turns along the $360^o$ longitude as well as an initial field strength of about $130\rm kG$ were necessary for a coherent radial rise of the tube, that the tilt angle was significantly influenced not only by the initial twist and Coriolis force but also by the local convection, and that meridional circulation was able to advect initially weak flux tubes toward the poles when they reached the top of the domain. They showed moreover a quantitative study of the effect of magnetic diffusivity by varying the Prandtl number in their simulations. \citet{Fan08} also investigated the effects of spherical geometry but on an $\Omega$-loop (a flux tube made buoyant in a limited range of longitudes) and thus focused on the properties of simulated individual active regions, but without convection. We here propose to compute the dynamical evolution of an $\Omega$-loop both in an isentropic and fully convective spherical shell, seeking to answer the following questions.

\begin{itemize}
\item[-] What are the physical processes at the origin of the observed tilt angles of active regions and is it possible to reproduce Joy's law?
\item[-] What is the influence of convective motions on a rising $\Omega$-loop and the subsequent bipolar emerging regions?
\item[-] Can we quantify the effects of large-scale mean flows?
\item[-] How important is the twist of the magnetic structures in the evolution within the convection zone and on the morphology of EFRs?
\item[-] How does the structure of simulated active regions compare to observations?
\end{itemize}

This work is thus an attempt to answer those questions using numerical simulations of a convective spherical shell, using the Anelastic Spherical Harmonic (ASH) code which solves the MHD equations under the anelastic approximation, valid within the bulk of the convection zone. We do not address here the evolution of emerging flux in the solar atmosphere which needs to be computed in a fully compressible simulation. Extensive studies of magnetic flux emergence in the atmosphere have been performed \citep[e.g.][]{Magara04, Manchester04, Murray06, Martinez08, Archontis08} which showed that magnetic flux reaching the photosphere could undergo a dynamical expansion into the atmosphere as a result of the non-linear growth of the magnetic buoyancy instability and that it is difficult for a twisted flux tube to rise bodily into a corona because of the heavy plasma trapped in the magnetic dips. Successive magnetic reconnections could however help to achieve actual emergence of twisted magnetic loops into the corona \citep{Pariat04}.

This article is organized as follows.
Section \ref{sect_model} presents the model chosen, the hydrodynamic background as well as the initial conditions which will favor the emergence of individual active regions. The following short Section \ref{sect_cases} presents the various cases calculated, which will be considered in an isentropic background (without convection) and with convective motions acting on the magnetic loop. Sections \ref{sect_isen} and \ref{sect_conv} respectively presents the results of the MHD simulations in the isentropic and convective cases. Finally, we discuss the results and conclude in Section \ref{sect_conclu}.

\section{THE MODEL}
\label{sect_model}
\subsection{Anelastic MHD Equations}

All calculations presented in this work were computed with the ASH code, which solves anelastic magnetohydrodynamic equations in a rotating spherical shell.ASH uses a
pseudo-spectral method in space and semi-implicit approach in time
\citep[e.g.,][]{Clune99,Miesch00,Brun04}. 
The full nonlinear evolution of the velocity and magnetic fields is computed, whereas the equations are linearized in
thermodynamic variables with respect to a spherically symmetric mean
state to have density $\bar{\rho}$, pressure $\bar{P}$, temperature
$\bar{T}$, and specific entropy $\bar{S}$. Perturbations are denoted as
$\rho$, $P$, $T$, and $S$. The equations being solved are

\begin{equation}
{\bf\nabla}\cdot(\bar{\rho}{\bf v})=0,
\end{equation}
 \begin{equation} {\bf\nabla}\cdot{\bf B}=0,
\end{equation}
\begin{eqnarray}
\bar{\rho}[\frac{\partial{\bf v}}{\partial t}&+&({\bf v}\cdot{\bf
    \nabla}){\bf v}+2\Omega_0\times {\bf v} ]=-{\bf \nabla} P
+\rho {\bf g}\\ \nonumber
&+&\frac{1}{4\pi}{\bf (\nabla \times B) \times B}
-{\bf \nabla \cdot \cal D}-[{\bf \nabla}\bar{P}-\bar{\rho}{\bf g}],
\label{eqNS}
\end{eqnarray}
\begin{eqnarray}
\bar{\rho}\bar{T}\frac{\partial S}{\partial t}&+&\bar{\rho}\bar{T}{\bf
  v}\cdot{\bf \nabla}(\bar{S}+S)={\bf
  \nabla}\cdot [\kappa_r\bar{\rho}c_p{\bf
    \nabla} (\bar{T}+T)\\ \nonumber
&+&\kappa_{0}\bar{\rho}\bar{T}{\bf
    \nabla}\bar{S}+\kappa \bar{\rho} \bar{T}{\bf \nabla} S ]+\frac{4\pi\eta}{c^2}{\bf j}^2 \\  \nonumber
&+&2\bar{\rho}\nu\left[e_{ij}e_{ij}-\frac{1}{3}({\bf \nabla \cdot
v)}^2\right],
\end{eqnarray}
\begin{equation}
\frac{\partial {\bf B}}{\partial t}={\bf \nabla \times} ({\bf v\times \bf B})-{\bf \nabla \times}(\eta {\bf \nabla \times B})
\end{equation} 

\noindent where ${\bf v}=(v_r,v_{\theta},v_{\phi})$ is the local
velocity in spherical coordinates in the frame rotating at a constant
angular velocity $\Omega_{0}$, ${\bf g}$ is the gravitational
acceleration, ${\bf B}=(B_r,B_{\theta},B_{\phi})$ is the magnetic
field, ${\bf j}=(c/4\pi)({\bf \nabla \times B})$ is the current
density, $c_p$ is the specific heat at constant pressure, $\kappa_r$
is the radiative diffusivity (fitted to a one-dimensional seismically calibrated solar model, see \citealt{Brun02a}), $\eta$ is the effective magnetic
diffusivity, and $\cal D$ is the viscous stress tensor. To represent the unresolved subgrid-scale processes, the code uses an effective eddy viscosity $\nu$ and an effective thermal eddy diffusivity. They are both here to model the enhanced diffusion associated with turbulent convective motions in the bulk of the convection zone. In the present simulations, they are chosen to vary in radius only as $\bar{\rho}^{-1/3}$. The thermal diffusion $\kappa_0$ acting on the mean
entropy gradient (contrary to the thermal eddy diffusivity $\kappa$ which acts on the entropy perturbation) occupies a narrow region in the upper convection
zone. Its purpose is to transport heat through the outer surface where
radial convective motions vanish \citep{Gilman81, Wong94} and is thus chosen such that the entire luminosity will be transported at the top of the domain by the so-called unresolved heat flux $F_u=-\kappa_0 \bar{\rho} \bar{T} d\bar{S}/dr$. The diffusivity $\kappa$ is purely dissipative and acts to smooth out entropy variations, whereas $\kappa_0$ is essentially a cooling term near the top of our computational domain. To complete the set of equations, we use the
linearized equation of state,

\begin{equation}
\frac{\rho}{\bar{\rho}}=\frac{P}{\bar{P}}-\frac{T}{\bar{T}}=\frac{P}{\gamma\bar{P}}-\frac{S}{c_p},
\end{equation}

\noindent where $\gamma$ is the adiabatic exponent (equal to $5/3$ here), and assume the ideal gas law

\begin{equation}
\bar{P}={\cal R} \bar{\rho}\bar{T},
\end{equation}

 \noindent where $\cal R$ is the ideal gas constant, taking into
  account the mean molecular weight $\mu$ corresponding to a mixture
  composed roughly of 3/4 of hydrogen and 1/4 of helium per mass. The reference or
mean state (indicated by overbars) is derived from a one-dimensional
solar structure model \citep{Brun02a}. It begins in hydrostatic balance so that the bracketed
term on the right-hand side of Equation\ref{eqNS} initially
vanishes. However, as the simulation evolves, turbulent and magnetic
pressures drive the reference state slightly away from strict hydrostatic balance.

The ASH code uses a spectral decomposition for all the variables, namely, spherical harmonics in the horizontal direction and Chebyshev polynomials in radius. The collocation points are thus the zeros of the Legendre polynomials in latitude, uniformly distributed in longitude, and the distribution is denser at both ends of the domain in the radial direction. The code integrates the MHD equations presented above over $0.72 R_\odot \leq r \leq 0.96 R_\odot$ (with an overall density contrast of about $24$ between top and bottom), $0 \leq \theta \leq \pi$ and $0 \leq \phi \leq 2\pi$. The resolution is chosen here such that at least $20$ points in radius and in latitude are contained inside the flux tube section. The minimal resolution used here is thus $N_r \times N_\theta \times N_\phi = 256 \times 512 \times 1024$.

Finally, the boundary conditions for the velocity are impenetrable and
stress-free at the top and bottom of the shell. We impose a constant
entropy gradient top and bottom for the isentropic case and for the
fully convective case, a latitudinal entropy gradient is
imposed at the bottom, as in \citet{Miesch06}.
In all cases, we match the magnetic field to an external potential
magnetic field at the top and the bottom of the shell \citep{Brun04}.

\subsection{Initial Conditions to Favor the Creation of $\Omega$-loops}
\label{sect_init}

In this paper, we implement initial conditions for the magnetic field similar to what was used in \citet{Jouve09}. An initially axisymmetric magnetic structure is embedded in an unmagnetized
stratified medium.  In order to keep a divergenceless magnetic field,
we use a toroidal-poloidal decomposition,

\begin{equation}
{\bf B}={\bf \nabla\times\nabla\times}(C {\bf e}_r) +{\bf
\nabla\times}(A {\bf e}_r)
\end{equation}

\noindent the expressions used for the potentials $A$ and $C$ for the flux tubes are:

\begin{equation}
A=-A_{0} \, r \, \exp\left[-\left(\frac{r-R_t}{a}\right)^2\right]
\times
\left[1+\tanh\left(2\frac{\theta-\theta_t}{a/R_t}\right)\right]
\end{equation}

\begin{equation}
C=-A_{0} \frac{a^2}{2} \, q \,
\exp\left[-\left(\frac{r-R_t}{a}\right)^2\right] \times
\left[1+\tanh\left(2\frac{\theta-\theta_t}{a/R_t}\right)\right]
\end{equation}

\noindent where $A_0$ is a measure of the initial field strength, $a$ is the tube radius,
 $(R_t,\theta_t)$ is the position of the tube center and $q$ is the
 twist parameter.

In \citet{Jouve09}, we derived an expression for the winding degree of the field
lines (i.e., the number of turns that the field lines make over the
whole tube length $2\pi R_t \sin \theta_t$, assuming a uniform twist): 

\begin{equation}
n=\frac{\pi R_t \sin\theta_t}{2 a}\tan\psi,
\end{equation}

where the tangent of the pitch angle $\psi$ is related to the twist parameter via the following expression, considering that we are at $\theta=\theta_t$ and at the tube periphery $r=R_t+a$, as in \citet{Jouve09}:

\begin{equation}
\tan\psi \approx \frac{q a}{R_t+a}
\end{equation}

In all cases, the tube radius is
set to $a=2\times10^9 \,\, \rm cm$, about a 10th of the depth of the
modeled convection zone and is introduced at the base of the CZ at
$R_t=5.2 \times 10^{10}\,\, \rm cm$. If we consider that the typical flux in an observed active region is $10^{21}-10^{22} \rm Mx$ and that an initial flux tube has a magnetic field of about $5\times 10^4 \rm G$ as in our simulations, the conservation of flux tells us that the radius of the loop should be in reality closer to $8 \times 10^7 \rm cm$. Our flux tubes are thus bigger than what could be expected. Different values for the loop radius have been shown to influence flux emergence \citep[see also the discussion of Section \ref{sect_conclu}]{DSilva93, Emonet98}. We here limit our parameter study to a fixed loop radius \citep[see][for a different set of parameters]{Jouve09}. The initial field strength $A_0$, the initial twist of the field lines $q$ as well as the latitude of introduction $\theta_t$ will be varied in our models to investigate the influence of these various parameters.

In order to get a flux tube buoyant on a small portion in longitude only, we initially apply a perturbation on the background entropy field located at the initial position of he flux tube in $(r,\theta)$ and 
possessing a Gaussian profile in longitude $\phi$. The entropy perturbation has thus the following expression:

\begin{eqnarray}
\label{eq_Spert}
S_{\rm {in}}-S_{\rm {ext}}&=&A_{S} \times \exp\left[-\left(\frac{r-R_t}{a}\right)^2\right] \times \\ \nonumber
& &\frac{2R_t}{a} \times \frac{1}{\cosh^2 \left(2\frac{\theta-\theta_t}{a/R_t}\right)}  \times\\ \nonumber
& & \left[-C_S+\exp \left( -(\frac{\phi-\phi_0}{\phi_{e}})^2\right) \right]
\end{eqnarray}

\noindent where $A_{S}$ is the amplitude of the entropy perturbation, $C_S$ is a constant value controlling the buoyancy of the rest of the loop, $\phi_0$ is the longitude of maximum perturbation (or buoyancy), and $\phi_{e}$ is the extent of the entropy perturbation.

The effect of such a perturbation on entropy is to produce an additional density deficit inside the flux tube at a particular location in longitude. As a consequence, we can derive the new maximum density contrast between the tube and its surroundings, which will result from this additional perturbation. Total pressure equilibrium is not enforced at the beginning of the simulation and since an initial magnetic tension is also present in most of our cases (the twisted cases), the loop is not initially in mechanical equilibrium outside the portion which is made buoyant, even if $C_S$ is modified. However, there is an initial adjustment visible at the very early stages of the dynamical evolution where the pressure perturbation inside the loop will tend to compensate for the magnetic pressure, in particular the loop gets slightly squeezed in an Alfv\'en time at the very beginning of the simulation, without much influence on the following evolution. We thus derive here only an indication for the efficiency of the magnetic buoyancy in a situation of pressure equilibrium and entropy perturbation (not exactly satisfied here) in writing the following relations respectively for the total pressure and the entropy:

\begin{equation}
P^{g}_{\rm{ext}}-P^{g}_{\rm{in}}=\frac{B^2}{8\pi}
\end{equation}

\begin{equation}
\rm{Max}\left(S_{\rm{in}}-S_{\rm{ext}}\right)=\Delta S=A_S\times\frac{2R_t}{a}\times (1-C_S),
\end{equation}

where $P^{g}_{\rm{ext}}$ and $P^{g}_{\rm{in}}$ are the gas pressure respectively outside and inside the tube and where the maximum of $S_{\rm{in}}-S_{\rm{ext}}$ is given by Equation \ref{eq_Spert} and taken at $r=R_t$, $\theta=\theta_t$, and $\phi=\phi_0$. We then use an additional relation to make the density  $\rho$ appear:

\begin{equation}
S=c_v \ln P - c_p \ln \rho,
\end{equation}

\noindent with $c_p$ the specific heat at constant pressure (equal here to $3.4 \times 10^8 \rm erg.K^{-1}.g^{-1}$), $c_v$ the specific heat at constant volume, and $\gamma=c_p/c_v = 5/3$ the adiabatic index.

These equalities lead to the following relation between entropy, pressure, and density inside and outside the flux tube:

\begin{equation}
\Delta S=\rm{Max}\left(S_{\rm{in}}-S_{\rm{ext}}\right)=\rm{Max}\left(c_v \ln \frac{P_{\rm{in}}}{P_{\rm{ext}}} - c_p \ln \frac{\rho_{\rm{in}}}{\rho_{\rm{ext}}}\right)  
\end{equation}

and thus we find an expression for the maximum buoyancy of our magnetic structures:

\begin{equation}
\frac{\Delta \rho}{\rho}=\rm{Max}\left(\frac{\rho_{\rm{ext}}-\rho_{\rm{in}}}{\rho_{\rm{ext}}}\right)=1-\exp\left(-\frac{\Delta S}{c_p}\right)\left(1-\frac{B_0^2}{8\pi P_{\rm{ext}}}\right)^{1/\gamma}\approx 1-\left(1-\frac{\Delta S}{c_p}\right)\left(1-\frac{B_0^2}{8\pi \gamma P_{\rm{ext}}}\right)
\label{eq_rho}
\end{equation}

where $B_0$ will be the maximum field strength on the tube axis. A first-order Taylor expansion has been done to give an approximate value of the buoyancy, considering that both $\Delta S/c_p$ and $B_0^2/8\pi P_{\rm{ext}}$ are equally small values. In the remaining of the paper, we will use the following parameters for the standard cases for the entropy perturbation: $C_S=0.35$, $\phi_{e}=15^o$, $\phi_0=100^o$ and the average external pressure used to estimate the maximum buoyancy will be $P_{\rm{ext}}=4\times 10^{13} \rm dynes$. We use a fixed value for $C_S$ since we do not seek to maintain a perfect mechanical equilibrium for the rest of the loop, otherwise the value of $C_S$ should depend on the magnetic field.

\subsection{The Background Hydrodynamical Models}
\label{sect_hydro}

The same hydrodynamical background was used in this work as in \citet{Jouve09}. Namely, two situations will be investigated. In the first, the magnetic structure is introduced in an isentropic layer with solid body rotation and without convection. In the second situation, the convective instability is triggered and differential rotation as well as a large-scale meridional flow are self-consistently created within the bulk of the convection zone. The competition between buoyant loops and convective motions and large-scale flows will in this case be studied and compared to the stably stratified situation. In particular, we focus on the shape of the developing loop, its asymmetries and the morphology of the emerging radial field it generates. 
Moreover, we focus in this study on the comparison to observations of large-scale EFRs, extrapolating the magnetic configurations we get from our simulations at the upper boundary (i.e., $0.96 R_{\odot}$) to what could happen higher up, closer to the photosphere. We specifically investigate the evolution of flux, tilt angle and spot morphology during emergence.

\section{CASES CALCULATED}
\label{sect_cases}

We investigate in this study seven different cases. All parameters used for each case are summarized in Table \ref{tab}. Four of the cases go by pair with the same buoyancy (but different field strength and entropy perturbation: Cases 1 \& 3 and 2 \& 4) and the last three are extreme cases (the weakest Cases 0a and 0b and strongest field Case 5). For Cases 1, 2, 3, and 4, the twist parameter was also varied. The cases with a right-handed (or Positive) twist (corresponding to $q=30$) will be called by the case number with the suffix TwP, and for the left-handed (or Negative) twist, we will add TwN. For example, the case where $B_0=5\times10^4 \rm G$, $\Delta S=3380$ (corresponding to a buoyancy $\Delta \rho/\rho$ of $1.1 \times 10^{-5}$ according to Equation \ref{eq_rho}) and which has a right-handed twist $q=30$ will be called Case 1TwP. 

Some additional cases were also computed: 

Case 1 with the loop introduced at the latitude of $60^o$: Case 1TwP60 and $15^o$: Case 1TwP15. 

Case 3 without twist: Case 3NoTw and with the twist at its threshold value (defined in the last paragraph of Section \ref{twist}): Case 3TwTh.

\begin{deluxetable}{cccccccccc}

\centering
\tablecolumns{8}
\tablewidth{0pc}
\tablecaption{Key Parameters of the Various Cases. The amplitude of the magnetic field is given in $\rm G$, latitude in degrees and entropy perturbation $\Delta S$ in $\rm erg.K^{-1}.g^{-1})$.}

\tablehead{
\colhead{}    & $B_{0}$           & $B_0^2/8\pi P_{\rm{ext}}$ & Twist $q$ & Latitude & $\Delta S$ & $\Delta S/c_p$     & Buoyancy $\Delta\rho/\rho$  } 

\startdata
Case0a        & $2\times10^4$     & $4\times10^{-7}$     & $30$      & $30$     & $1690$     & $5\times10^{-6}$   & $5.1\times 10^{-6}$      \\ \hline 
Case0b        & $4\times10^4$     & $1.6\times10^{-6}$   & $30$      & $30$     & $1690$     & $5\times10^{-6}$   & $5.9\times 10^{-6}$      \\ \hline 
Case 1TwP     & $5 \times 10^4$   & $2.5\times10^{-6}$   & $30$      & $30$     & $3380$     & $9.9\times10^{-6}$ & $1.1 \times 10^{-5}$     \\ \hline 
Case 1TwP15   & $5 \times 10^4$   & $2.5\times10^{-6}$   & $30$      & $15$     & $3380$     & $9.9\times10^{-6}$ & $10^{-5}$                \\ \hline
Case 1TwP60   & $5 \times 10^4$   & $2.5\times10^{-6}$   & $30$      & $60$     & $3380$     & $9.9\times10^{-6}$ & $10^{-5}$                \\ \hline
Case 1TwN     & $5 \times 10^4$   & $2.5\times10^{-6}$   & $-30$     & $30$     & $3380$     & $9.9\times10^{-6}$ & $10^{-5}$                \\ \hline 
Case 2TwP     & $10^5$            & $9.9\times10^{-6}$   & $30$      & $30$     & $3380$     & $9.9\times10^{-6}$ & $1.6\times 10^{-5}$      \\ \hline 
Case 2TwP15   & $10^5$            & $9.9\times10^{-6}$   & $30$      & $15$     & $3380$     & $9.9\times10^{-6}$ & $1.6\times 10^{-5}$      \\ \hline 
Case 2TwP60   & $10^5$            & $9.9\times10^{-6}$   & $30$      & $60$     & $3380$     & $9.9\times10^{-6}$ & $1.6\times 10^{-5}$      \\ \hline 
Case 2TwN     & $10^5$            & $9.9\times10^{-6}$   & $-30$     & $30$     & $3380$     & $9.9\times10^{-6}$ & $1.6\times 10^{-5}$      \\ \hline 
Case 3TwP     & $10^5$            & $9.9\times10^{-6}$   & $30$      & $30$     & $1690$     & $5\times10^{-6}$   & $10^{-5}$                \\ \hline 
Case 3NoTw    & $10^5$            & $9.9\times10^{-6}$   & $0$       & $30$     & $1690$     & $5\times10^{-6}$   & $1.1 \times 10^{-5}$     \\ \hline 
Case 3TwTh    & $10^5$            & $9.9\times10^{-6}$   & $10$      & $30$     & $1690$     & $5\times10^{-6}$   & $1.1 \times10^{-5}$      \\ \hline 
Case 3TwN     & $10^5$            & $9.9\times10^{-6}$   & $-30$     & $30$     & $1690$     & $5\times10^{-6}$   & $1.1 \times 10^{-5}$     \\ \hline 
Case 4TwP     & $1.35 \times 10^5$& $1.8\times10^{-5}$   & $30$      & $30$     & $1690$     & $5\times10^{-6}$   & $1.6\times10^{-5}$       \\ \hline 
Case 5TwP     & $2\times 10^5$    & $4.\times10^{-5}$    & $30$      & $30$     & $3380$     & $9.9\times10^{-6}$ & $3.4\times10^{-5}$       
\enddata
\label{tab}
\end{deluxetable}

The values of the entropy perturbation were chosen to ensure a rise time of the same order of the convective turnover time, i.e., between 10 and 40 days approximately. We can compare the buoyancy of the loops with and without the entropy perturbation by setting $\Delta S=0$ in Equation \ref{eq_rho}. We find that the value of $\Delta\rho/\rho$ is multiplied by $7.65$ when the entropy perturbation is introduced in Case 1 (compared to the case without entropy perturbation) and by a factor $1.41$ in Case 5, so that the stronger field cases are less influenced by the entropy perturbation applied. Moreover, the values of the entropy perturbations used here are comparable in amplitude to the typical fluctuations measured at the base of our computational domain in our convective runs.

\section{EVOLUTION OF A SINGLE LOOP WITHOUT CONVECTION}
\label{sect_isen}

We first compute the evolution of a single $\Omega$-loop evolving in a stable background with respect to convection. It will first confirm that we indeed correctly control the buoyancy of the flux tube with the initial conditions we apply.
Secondly, it will help us see if the important parameters for the evolution of a uniformly buoyant flux tube are still the same here and what differences may be pointed out. 

\subsection{Favorable Conditions for Emergence}

\subsubsection{Twist}
\label{twist}

We know from previous experiments that a certain amount of twist is needed for a magnetic flux tube to be able to rise coherently in a convectively stable (or unstable) background \citep[e.g.,][]{Emonet98}. Indeed, in a non-twisted case, the azimuthal vorticity created by the gravitational torque acting on the loop cannot be compensated. Two counter vortices thus quickly appear and break the loop apart, preventing it from emerging at the surface. It was argued previously that this situation may differ when an $\Omega$-loop is considered, instead of a uniformly buoyant structure \citep{Abbett00}. The additional magnetic tension coming from the deformation of the magnetic torus may oppose the vorticity generation. 

\begin{figure}[h!]
	\centering
	\includegraphics[width=7.5cm]{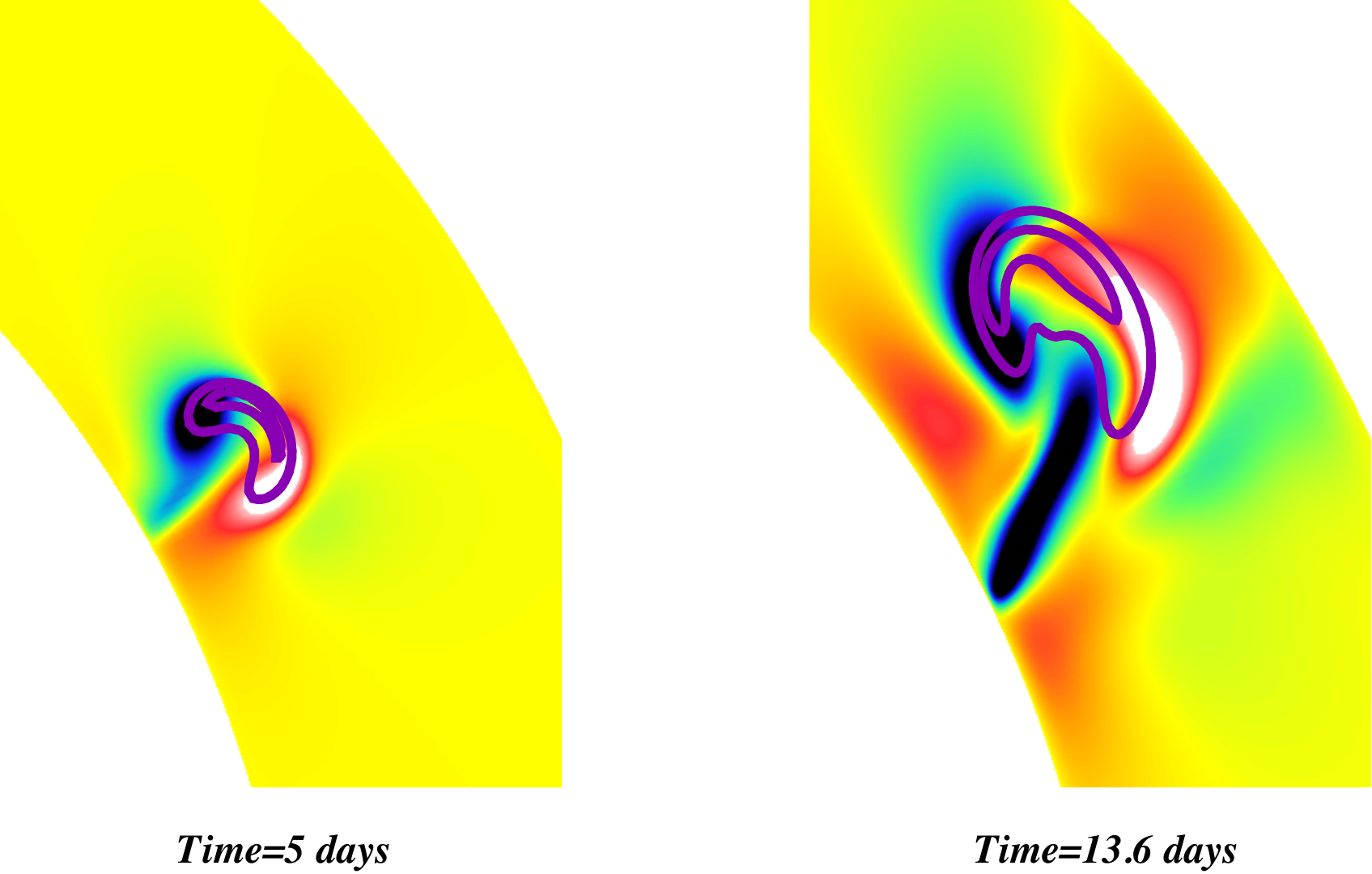}
	\caption{Cross section of the untwisted $\Omega$-loop of Case 3NoTw, introduced at a latitude of $30^o$, at two instants in the evolution, $5$ and $13.6$ days. Magenta contours of the azimuthal component of the magnetic field are superimposed to the azimuthal vorticity which presents a characteristic profile of two counter vortices.}
	\label{figure_notw}
\end{figure}

Figure \ref{figure_notw} shows the section of an untwisted $\Omega$-loop corresponding to Case 3NoTw of Table \ref{tab}. The zoom is made on the longitude where the loop is caused to be buoyant, thanks to the combined effects of magnetic pressure and entropy perturbation. Two instants in the evolution are shown. We represent here the contours of the azimuthal (or axial) component of the magnetic field superimposed on the azimuthal vorticity. It is rather clear in this case (especially on the left panel) that strong vortices are formed inside the flux tube, which have the tendency to significantly deform the magnetic structure. However, as the loop rises further up, it becomes more arched and more magnetic tension is generated. As a consequence, even without any twist, some coherence is maintained in this case, even if a large amount of flux is lost in the process.

\begin{figure}[h!]
	\centering
	\includegraphics[width=7.5cm]{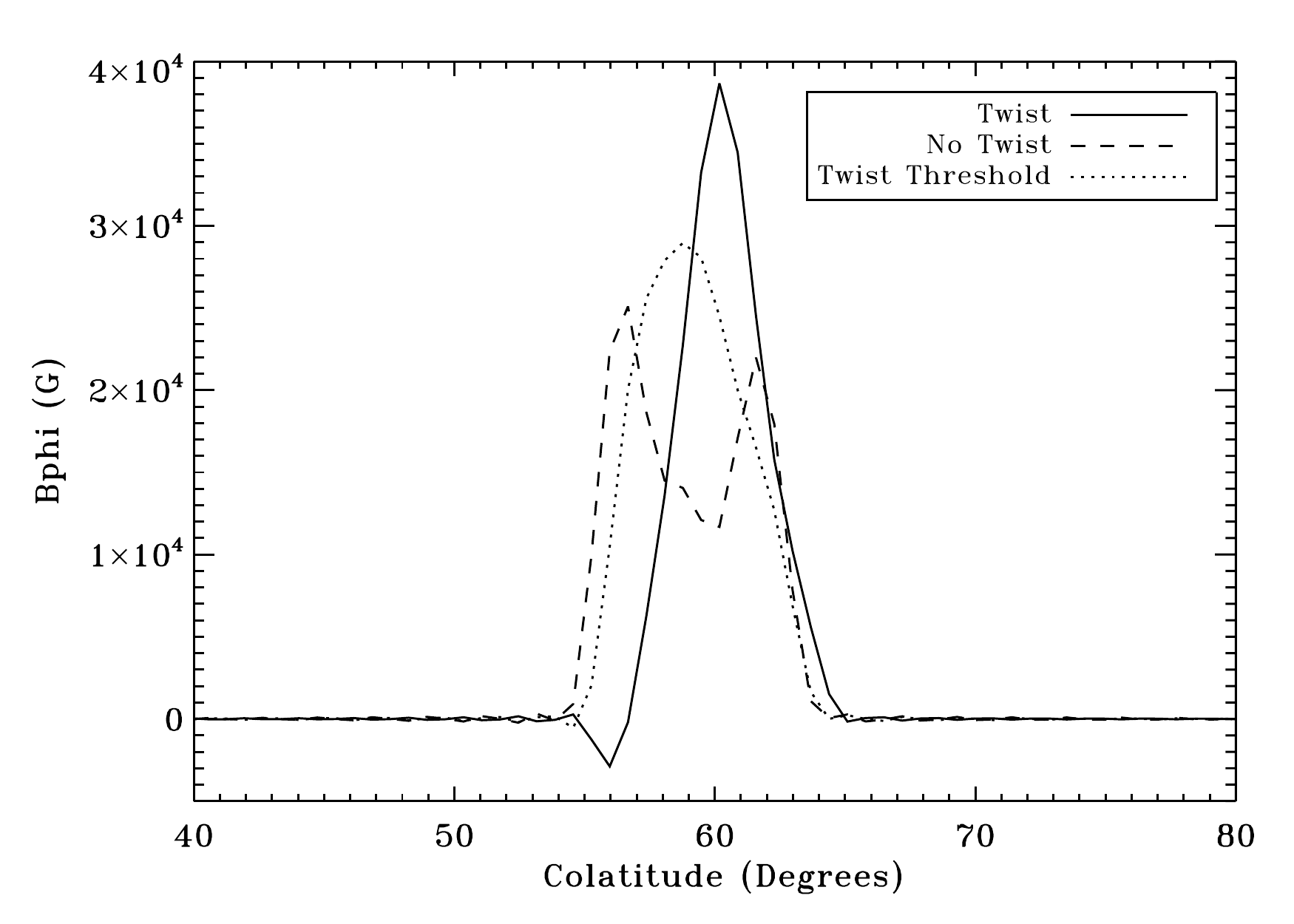}
	\caption{Profile of the toroidal field at the radius where it peaks after 6 days of evolution (i.e., $0.82 R_{\odot}$). The three cases represented are the untwisted case ($q=0$) Case 3NoTw, the twisted case Case 3TwP ($q=30$), and the intermediate case Case 3TwTh where the twist value is at its threshold ($q=10$).}
	\label{figure_notw2}
\end{figure}

This flux loss is visible in Figure \ref{figure_notw2} where the profile of the toroidal magnetic field is shown at the radius where it peaks after about $6$ days of evolution, as a function of colatitude. We remind that the flux ropes were introduced here at $60^o$ of colatitude and with an amplitude of $10^5 \rm G$. In the non-twisted case (dashed line on the figure), we clearly see that the two counter vortices have created two distinct magnetic flux concentrations, whose maximum amplitude reaches a value of $2.5 \times 10^4 \rm G$ while the maximum amplitude of the twisted case is $4\times 10^4 \rm G$. The flux rope has been eroded by the vorticity generation and the two vortices tend to dominate the further evolution of the magnetic structure which separates into two flux concentrations. However, along the loop (following its axis), this double-peak structure is not as clearly visible as at the longitude we chose here. This feature may be the consequence of the asymmetry which develops between the leading and trailing leg. We will come back on this in the next section.
 
We decided to define the twist threshold as the value above which the double-peak structure is not present anymore at any longitude. With this criterion, we find that the twist threshold for this case is $q=10$, corresponding to $0.54$ turns along the $15^o$ of extension of the $\Omega$-loop. We note that this threshold value is reduced by a factor $1.8$ compared to the uniformly buoyant flux tubes of \citet{Jouve09}. This is due to the additional tension coming from the arching of the $\Omega$-loop in this case, which has already the effect of limiting the vorticity generation inside the magnetic structure. This result is in agreement with the work of \citet{Abbett00}. The case where the twist is set at its threshold value is also represented in Figure \ref{figure_notw2} (dotted line). In this case, the maximum value of the field is reduced compared to the more twisted case but the coherence of the loop is kept.

\subsubsection{Buoyancy versus Rotation}

The buoyancy force is proportional to $\Delta \rho$ which is given by expression (\ref{eq_rho}). While the loop rises because of the buoyancy force, a retrograde longitudinal flow develops in order to conserve angular momentum (see section below). As a consequence, a Coriolis force oriented toward the rotation axis and perpendicular to this axis also develops, opposing the component of the buoyancy force perpendicular to the rotation axis. The only uncompensated force is then parallel to the rotation axis and oriented toward the pole. If the buoyancy force is not strong enough, the loop will thus have the tendency to rise parallel to the rotation axis and emerge at a much higher latitude than where it was introduced. This is the case at fixed $\Delta S$ when the magnetic energy is not strong enough. We recover this situation when the poloidal field is too weak (corresponding to a low twist value, enough to maintain coherence of the loop but not enough to make it rise radially).  

\begin{figure}[h!]
	\centering
	   \includegraphics[width=7.5cm]{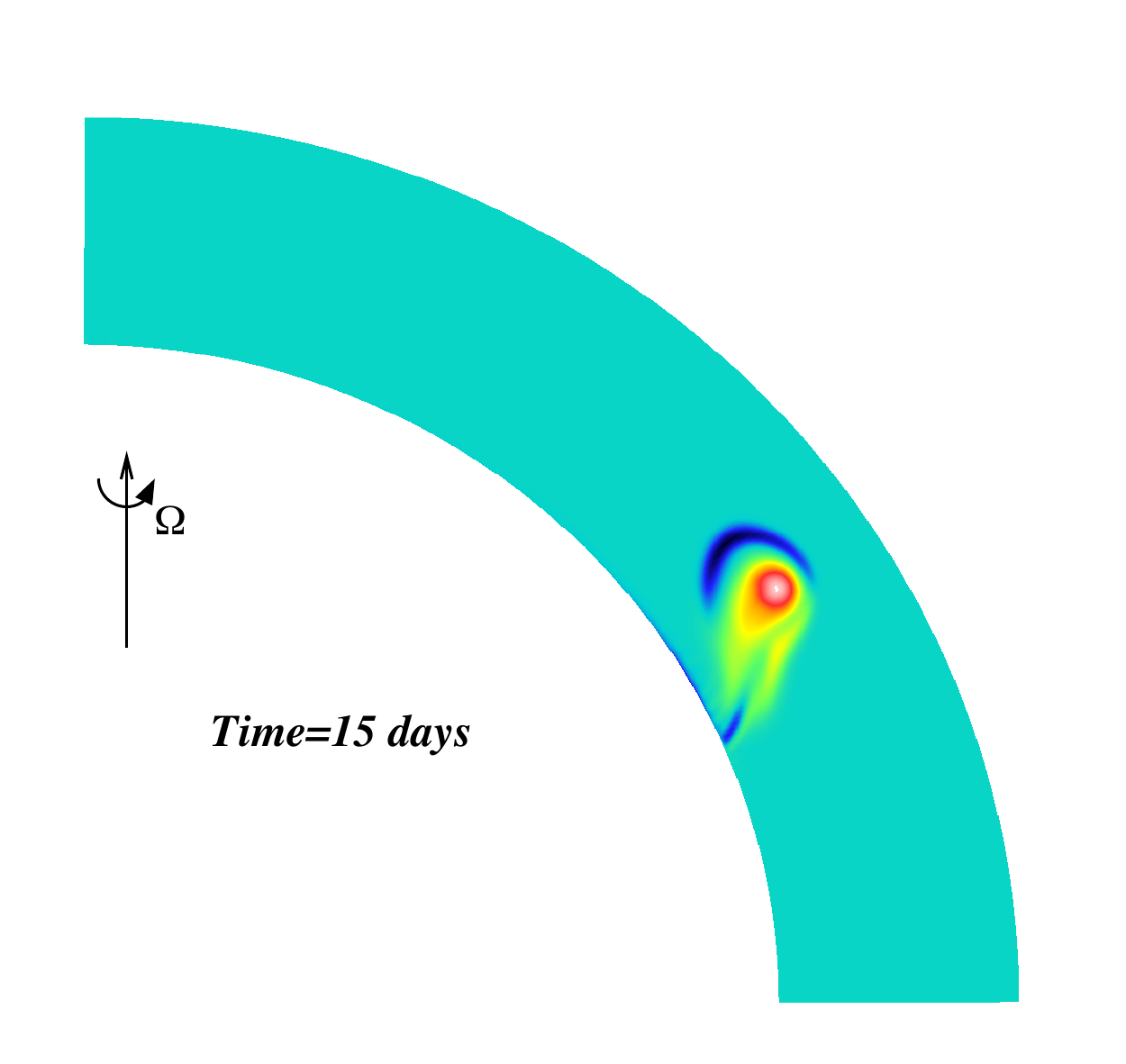}
\caption{Section of $B_\phi$ at $t=15$ days at the longitude of maximum buoyancy for Case 0a. The trajectory of the loop is here clearly parallel to the rotation axis.}
	\label{figure_vertical}
\end{figure}

Figure \ref{figure_vertical} is an illustration of a case where the rise is mainly parallel to the rotation axis. In this Case 0a, the twist was set to $q=30$ but the initial axial field strength was set to $2\times10^4 \rm G$, and the entropy perturbation to the  rather weak value of $\Delta_S=1680$, corresponding to a density perturbation of $\Delta\rho/\rho=5\times10^{-6}$. In this case, after $15$ days of evolution, the trajectory of the loop is clearly parallel to the rotation axis, since the Coriolis force has strongly acted on the magnetic field. Indeed, as shown by \citet{Choudhuri87} in the framework of the thin flux tube approximation, while the loop rises, it develops a retrograde longitudinal flow in its interior. A Coriolis force is associated with this longitudinal flow which tends to oppose the outward component of the buoyancy force perpendicular to the rotation axis, leaving only the component parallel to the rotation axis to act on the ring. Interestingly, \citet{Choudhuri87} find that in the adiabatic cases (similar to what we are studying in this section), the transition between vertical and radial rise happens between $\Delta \rho/\rho=2\times 10^{-6}$ and $\Delta \rho/\rho=2\times 10^{-5}$. We also find here that Cases 0a and 0b, where $\Delta \rho/\rho$ is around $5-6\times 10^{-6}$, tend to be strongly influenced by the Coriolis force while the stronger cases, where  $\Delta \rho/\rho \geq 10^{-5}$, exhibit a much more radial trajectory. 

We note that due to the low value of the buoyancy used for Case 0a, the rise of the loop is rather slow (about 40 days are needed to reach the top of our domain). In order to ensure an evolution dominated by the dynamical effects and not only by diffusion, we reduced the magnetic diffusivity from $7.95 \times 10^{11} \rm cm^2.s^{-1}$ to $1.59 \times 10^{11} \rm$, and thus used a magnetic Prandtl of 5 instead of 1 as with all the other cases. It does not change the fact that the rise is parallel to the rotation axis in this case, it is just to compute a case where the rise time is significantly lower than the diffusive time (which would be of order $R^2/\eta=58$ days for the $Pm=1$ case and thus too close to the rise time of $40$ days).

The same kind of evolution parallel to the axis of rotation is found when stronger tubes are introduced with a lower twist. Indeed, if the twist is low as we saw before, the amplitude of the magnetic field inside the loop quickly decreases until it reaches a value where the buoyancy force is not strong enough to keep the radial trajectory. This is visible on the right panel of Figure \ref{figure_notw} where the initial axial field was $10^5 \rm G$ but the initial poloidal field was $0$.
To correctly define a minimum value for a radial rise here, it is thus necessary to take into account the three components of the magnetic field rather than focus on the axial field. Let us consider a case where the entropy perturbation is the weakest, corresponding to a slow rise and then to a maximum effect of the Coriolis force. We then find that for a twist parameter $q=30$ (about 1 turn over the extension chosen here), the initial magnetic field needed for a radial rise should be at least of $5\times 10^4 \rm G$. It can be argued that this value is strong compared to what could be expected at the base of the convection zone from dynamo action but it could be reduced if a stronger entropy perturbation was considered, which is locally possible in the turbulent environment existing at the base of the convection zone.\\

To summarize the two previous subsections, we can say that to maintain a coherent radial rise of our $\Omega$-loop in our simulation, an initial twist of $0.54$ turns along the extension of the loop, as well as an initial field strength of $50 \rm kG$ are necessary. We note that those two values are reduced compared to the uniformly buoyant calculations. This is promising since the observed twist values are of the order of less than 1 turn across the active region \citep{Chae05, Demoulin02} and typical fields of more than $50 \rm kG$ are difficult to produce in present dynamo calculations \citep{Brun04}.

\subsection{Asymmetry between the Trailing and Leading Legs}

\begin{figure*}[h!]
	\centering
	\includegraphics[height=12cm]{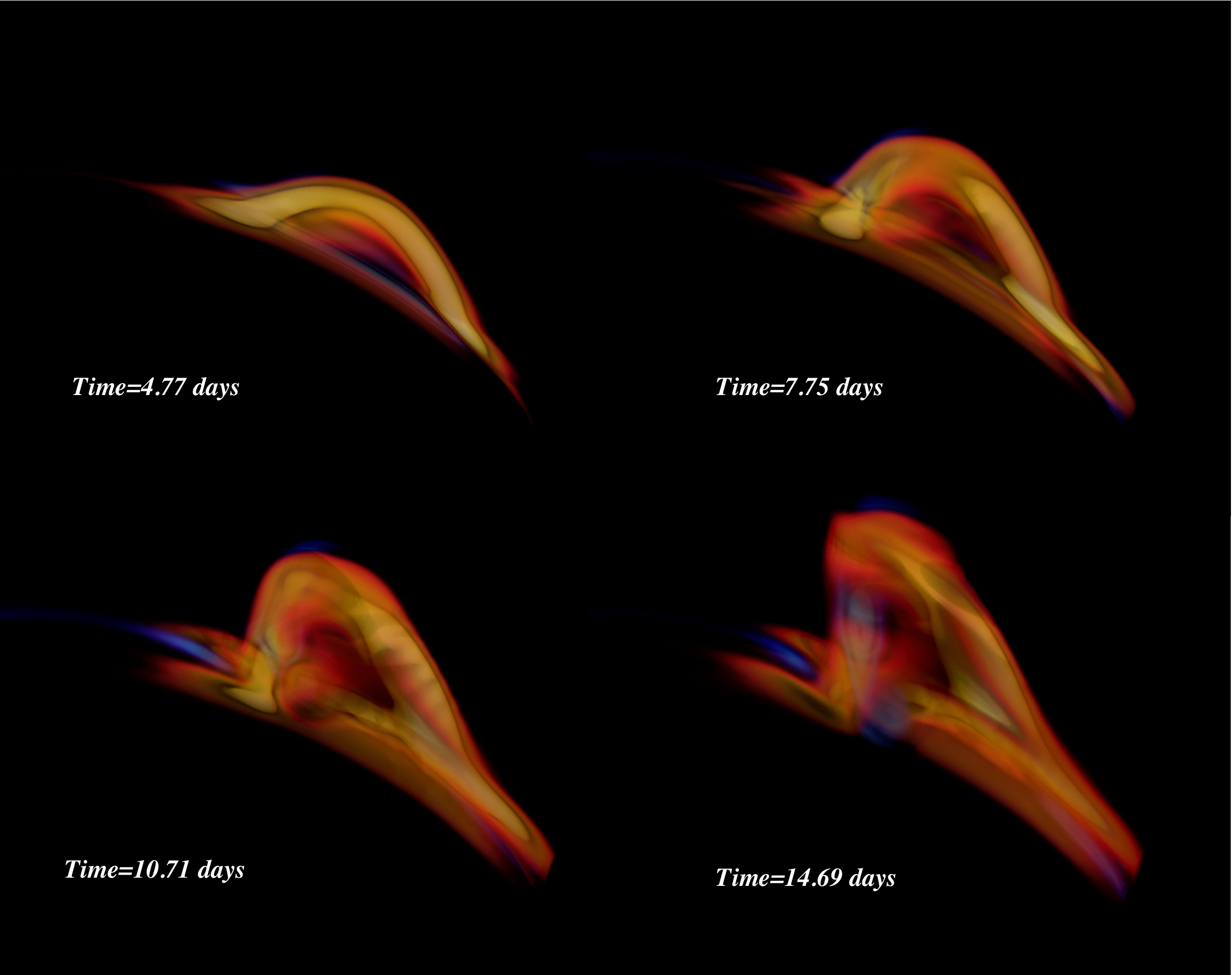}
	\caption{Volume rendering of $B_{\phi}$ while the loop rises through the isentropic layer in Case 1TwP. The loop is viewed up to the polar axis, from the South Pole. The asymmetry between the legs of the loop are obvious in these snapshots. The visualization was performed with \emph{SDvision@cea} \citep{Pomarede10}.}
	\label{figure_isen3D}
\end{figure*}

We now study the evolution of a sufficiently twisted and strong loop so that it will maintain its coherence while it rises radially. The loop is introduced at $30^o$ in latitude, with a maximum buoyancy (i.e., entropy perturbation) located at $100^o$ in longitude, with a twist parameter $q=30$ representing about $39$ turns around the tube axis over the whole $360^o$ and thus corresponding to about $1.6$ turns over the extension of the buoyant part ($\phi_{e}=15^o$). With this twist parameter (as we saw before), the loop rises radially for a field strength of $5\times 10^4 \rm G$. Figure \ref{figure_isen3D} shows a volume rendering of the toroidal field evolution in Case 1TwP.

The asymmetry between the trailing and leading legs develops during the evolution and is clearly seen in the last two panels of the figure, at $t=10.7$ days and $t=14.7$ days. This section will be devoted to the study of this asymmetry and the influence of buoyancy and field strength. The action of rotation and the associated Coriolis force over the whole loop will also be detailed here. This process will lead to the appearance of a tilt of the axis with respect to the east-west direction and will also depend on the magnetic field strength, as we will see in the following section.

Let us enter now into the details of the asymmetry between the trailing and leading legs and the effect of the field strength on it. This asymmetry, already seen in thin flux tubes calculations \citep[see][]{Caligari95}, is due to the difference in azimuthal velocity between the top of the loop and its feet. Indeed, since the flux tube needs to conserve its angular momentum $r \sin\theta \left(r\sin\theta \Omega + v_{\phi}\right)$, a retrograde flow with respect to the rotating frame will develop inside the loop while it rises (i.e., while $r$ increases). Since the apex of the loop is higher than the feet which stay attached to the bottom of the convection zone, a stronger retrograde flow appears at the apex. The profile of the flow along the magnetic structure then leads to the inclination asymmetry of the trailing and leading legs. In some of our cases, since the parameter $C_S$ is always kept the same, parts of the loop located outside the buoyant portion can be slightly denser than their surroundings. As a consequence, those regions may sink toward the base of the convection zone and favor the difference in azimuthal velocity causing the asymmetry. However, in most cases (except Cases 0a, 0b, 1TwP, and 1TwN), the magnetic buoyancy can overcome this effect and the rest of the loop is slightly positively buoyant, thus on the contrary reducing the asymmetry.

\begin{figure*}[h!]
	\centering
	\includegraphics[width=15cm]{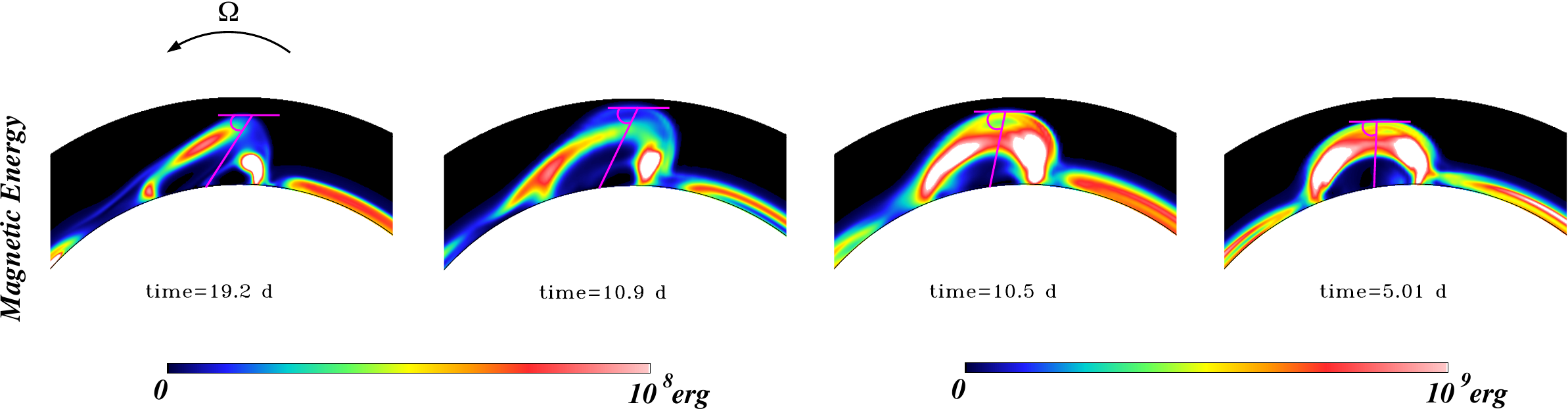}
	\includegraphics[width=15cm]{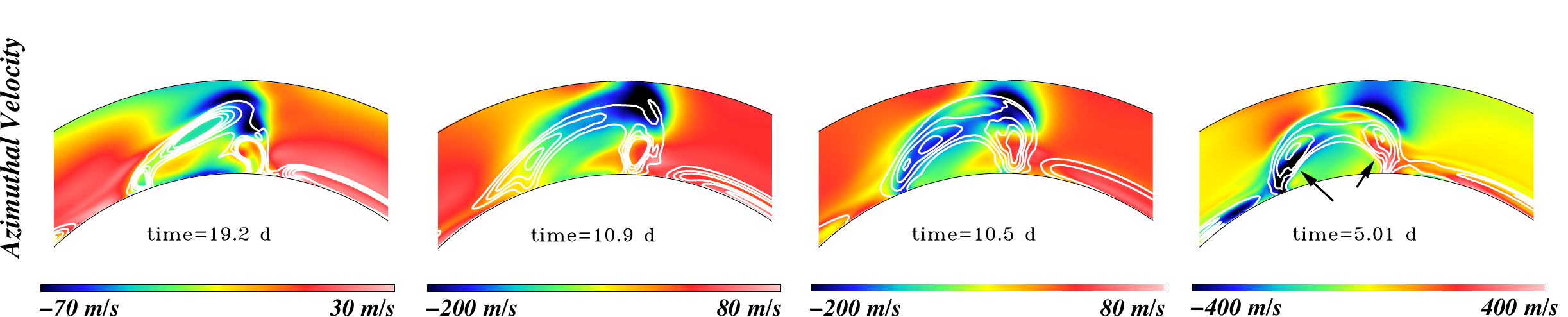}
	\caption{Cut at the latitude of $30^o$ of the magnetic energy (upper panel) and azimuthal velocity together with the white contours of magnetic energy (lower panel) for four different cases, Cases 0b, 1TwP, 2TwP, and 5TwP, when the $\Omega$-loops have approximately reached the same height $r=0.93 R_{\odot}$. The loops are viewed down from the North Pole. The color scale used for the magnetic energy is the same for the first two panels and the same for the last two.}
	\label{figure_eqsl}
\end{figure*}

However, this explanation is valid as long as the Lorentz forces do not dominate the flow evolution inside the loop. Figure \ref{figure_eqsl} shows a snapshot of the magnetic energy and the azimuthal velocity for four different cases: Cases 0b, 1TwP, 2TwP, and 5TwP. We note here that Case 0b has a magnetic field strength of $4\times 10^4 \rm G$ and is thus still significantly deflected toward the pole but is an interesting case because it maximizes the asymmetry between the legs of the loop. The snapshots were chosen so that all loops have reached the same height (namely $0.93 R_\odot$), which corresponds to different physical times since the buoyancy force (i.e., the $\Delta \rho$) is different in all cases. In order to quantify the asymmetry, we measure the angle between the tangent at the apex and the line going through the middle of the two footpoints and the apex. Those lines are shown in magenta in the top panels of Figure \ref{figure_eqsl}. We find that the angles have the following values: $45^o$, $50^o$, $75^o$, and almost $90^o$ (meaning no asymmetry in this Case 5TwP where the magnetic field and the buoyancy force are the strongest). We thus note that the asymmetry strongly decreases from Case 1TwP to Case 2TwP, even if the density deficits are initially not very different. The reason for this jump in the asymmetry is the difference in field strength ($10^5 \rm G$ for Case 2TwP against $5\times 10^4\rm G$ for Case 1TwP). Indeed, in this case, the Lorentz force starts to become important in the generation of azimuthal velocity: strong values of $v_\phi$ appear at the location of strong gradients of magnetic field. This is much more obvious in the stronger Case 5TwP. The profile of the azimuthal velocity shown on the last snapshot of the lower panel is clearly related to the gradients of magnetic field and the retrograde flow due to the conservation of angular momentum is so weak compared to the magnetic effects that it is not visible anymore. Instead, an azimuthal velocity directed toward the bisector of the tube appears on both sides of the loop (indicated by black arrows in the last panel), resulting in a symmetric arched structure (visible in the top right snapshot) with both feet showing an almost vertical orientation. 

If we write down the expression of the Lorentz force involved in the evolution of $v_\phi$, we have:

\begin{eqnarray}
\label{eq_vphi}
\frac{Dv_\phi}{Dt}&=&\frac{J_r B_\theta - J_\theta B_r}{4\pi} + \mbox{non-magnetic terms}\\ \nonumber
&=&\frac{B_\theta}{r\sin\theta}\left(\frac{\partial}{\partial\theta}\left(\sin\theta B_\phi\right)-\frac{\partial B_\theta}{\partial\phi}\right)\\ \nonumber
&-& \frac{B_r}{r}\left(\frac{1}{\sin\theta}\frac{\partial B_r}{\partial\phi}-\frac{\partial}{\partial r}(r B_\phi)\right) + \mbox{non-magnetic terms}
\end{eqnarray}

This equation shows that the regions of strong magnetic field gradients will correspond to the regions of production of $v_\phi$, which can be verified from the last panel of Figure \ref{figure_eqsl}, where the concentrations of $v_\phi$ are located mainly at the footpoints (where the gradients of $B_r$ and $B_\theta$ with respect to $\phi$ are strong) and higher up at the loop periphery (where the gradients of $B_\phi$ with respect to $r$ and $\theta$ are strong).

\begin{figure*}[h!]
	\centering
	\includegraphics[width=15cm]{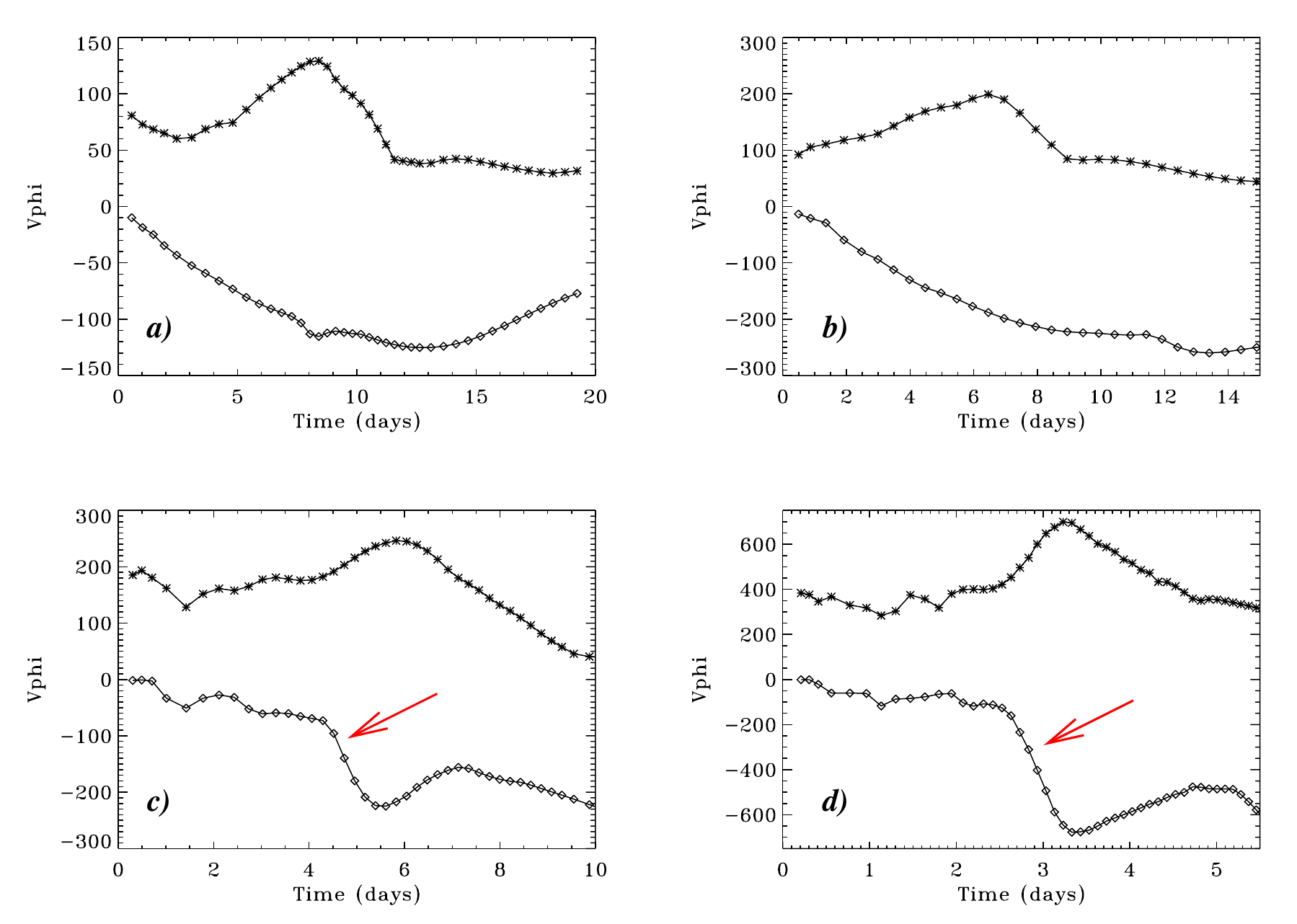}
	\caption{Evolution of the minimum and the maximum azimuthal velocity in the whole domain in the four cases, Case 0b (panel ($a$)), Case 1TwP (panel ($b$)), Case 2TwP (panel ($c$)), and Case 5TwP (panel ($d$)) of Figure \ref{figure_eqsl}.}
	\label{figure_eqsl2}
\end{figure*}

To be more precise, we can follow the evolution of the maximum and minimum of the azimuthal velocity which develops during the rise of the flux tube from the base of the convection zone toward the surface. In Figure \ref{figure_eqsl2}, we plot the maximum and the minimum of $v_\phi$, in order to understand the differences in the asymmetries seen in Figure \ref{figure_eqsl}. For each case, the maximum and minimum azimuthal velocities are measured over the whole domain but those extreme values are always reached inside or in the vicinity of the magnetic structures. Moreover, the various cases (except for Case 0b) only differ by the initial magnetic field strength of the loop and the differences in $v_{\phi}$ will thus be directly related to the differences in the magnetic properties. The most striking feature is the strong increase of the maximum velocity in all cases. This sharp increase happens in all cases when the loop reaches the mid-CZ, i.e., when the loop starts to get a significant curvature. If we look at Equation (\ref{eq_vphi}), we can understand that this sharp increase is due to the gradients with respect to the $\phi$-coordinate of $B_r$ and $B_\theta$ which appear only when the loop becomes arched, i.e., after a certain rise time, contrary to the gradients of $B_\phi$ with respect to $r$ and $\theta$, which exist at the beginning of the simulation already. If we now turn to look at the minimum of $v_\phi$, we realize that this change of slope is also visible at the same time but only for the strong field cases. They are indicated by the red arrows in panels ($c$) and ($d$). For the retrograde flow with respect to the rotating frame indeed, another ingredient plays a role, the conservation of angular momentum. As we said before, it will have a tendency to produce a negative $v_\phi$ which is dominant in the weak-field cases (panels ($a$) and ($b$)) but dominated by the Lorentz force terms in the strong field cases (panels ($c$) and ($d$)). As a consequence, strong gradients of the magnetic field and the associated magnetic tension will have the tendency to oppose the deformation of the loop while it rises and it explains why the asymmetry between the two legs of the loop is reduced when the initial field is strong enough (more than $10^5 \rm G$). This feature was already observed in thin-flux tube calculations \citep[e.g.,][]{Moreno94,Caligari95} and a similar argument involving the stiffness of intense fields was used to explain the weaker asymmetry seen in strong field cases. We note moreover that since the rise time of strong tubes is smaller, the Coriolis force will have less time to act on the loop and the asymmetry will also be reduced because of shorter rise times.

Going back to Figure \ref{figure_isen3D}, we see that when the loop forms and rises, some negative toroidal field is created (blue colors in the figures), while the initial toroidal field was purely positive. This can be explained by a local $\Omega$-effect: the azimuthal velocity gradient built at the periphery of the loop while it rises will shear the small poloidal field component present at the tube boundary mainly. This $\Omega$-effect, related to a shearing of poloidal field is one of the main ingredients of the dynamo mechanism thought to be responsible for the generation of magnetic field in differentially rotating astrophysical objects (see for example \citealt{Brun04a, Charbo05, Miesch05} for a review on these processes in the solar context). The presence of such an effect could be surprising, especially in the untwisted case without convection, since no poloidal field nor azimuthal velocity gradients exist at the initial stage. However, while the loop forms, the magnetic field is no longer purely toroidal since the field lines get curved and will thus get a mainly radial component. At the same time, gradients of azimuthal velocity will be created as shown above. Both these effects are then favorable to get an $\Omega$-effect which will be able to create additional toroidal field of both signs.

\subsection{Structure of the Emerging Regions}

We now turn to investigate the characteristics of the individual bipolar regions emerging at the top of our computational domain in all situations. In particular, we focus on the tilt angle, the asymmetry in the field strength in the two spots and the morphological properties of the regions which can be related to observations.

\subsubsection{Evolution of the Tilt Angle: Effects of Twist and Latitude}

We decide here to focus on four cases: Cases 1TwP and 1TwN and Cases 3TwP and 3TwN. The main difference between the two cases with right-handed twist is the field strength which is multiplied by 2 in Case 3 but buoyancy is kept the same through a change of initial entropy perturbation. Moreover, we know from previous studies \citep[e.g.,][]{Fan08} that the sign of the initial twist will have a dramatic effect on the tilt angle of the emerging region. We address here this question by computing Cases 1 and 3 with a right-handed as well as left-handed twist.  

\begin{figure*}[h!]
	\centering
	\includegraphics[height=12cm]{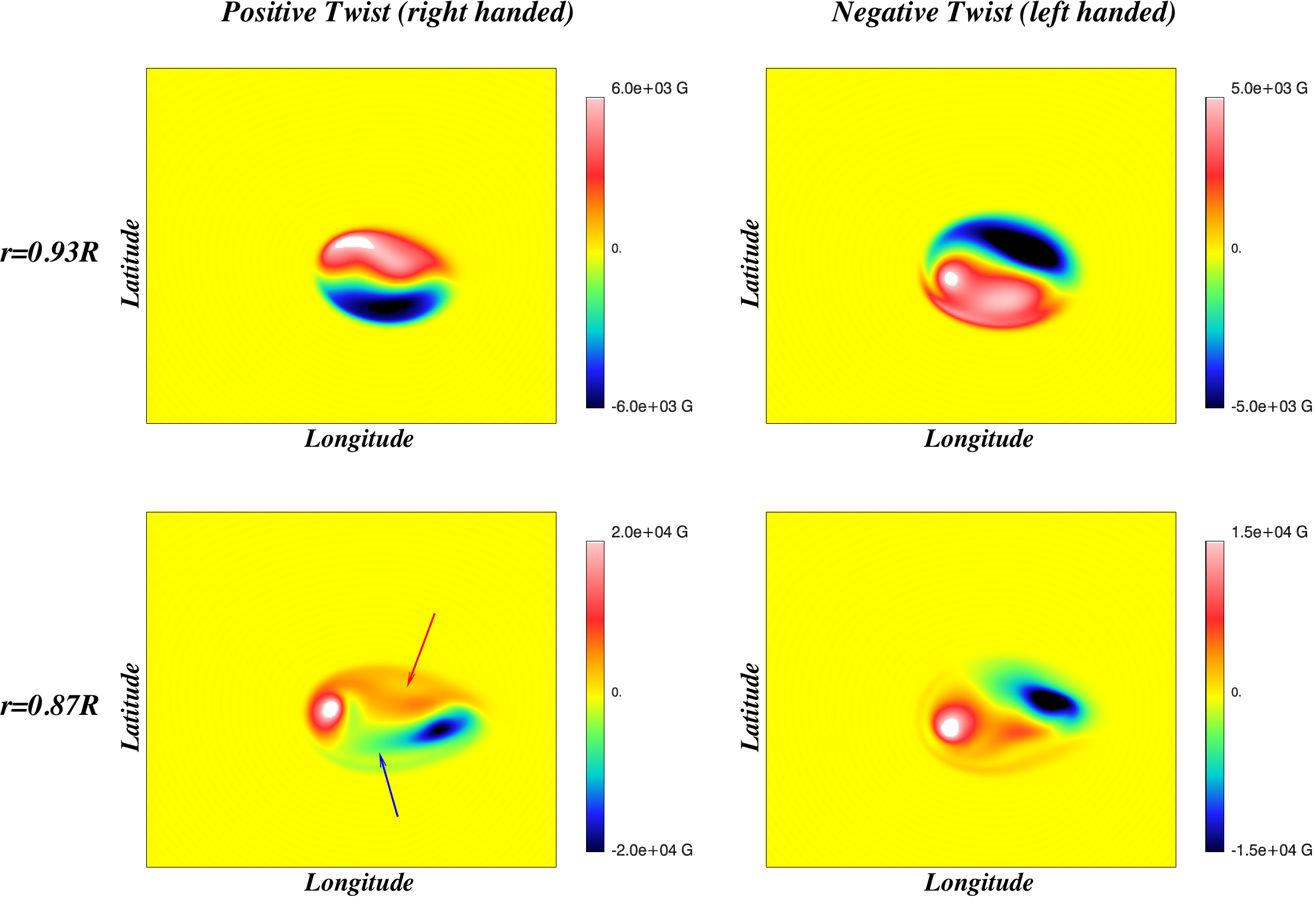}
	\caption{Zoom on the emerging radial field at $r=0.93R_\odot$ for Cases 3TwP and 3TwN where $B_0=10^5 \rm G$.}
	\label{figure_shsl0}
\end{figure*}

\begin{figure*}[h!]
	\centering
	\includegraphics[height=12cm]{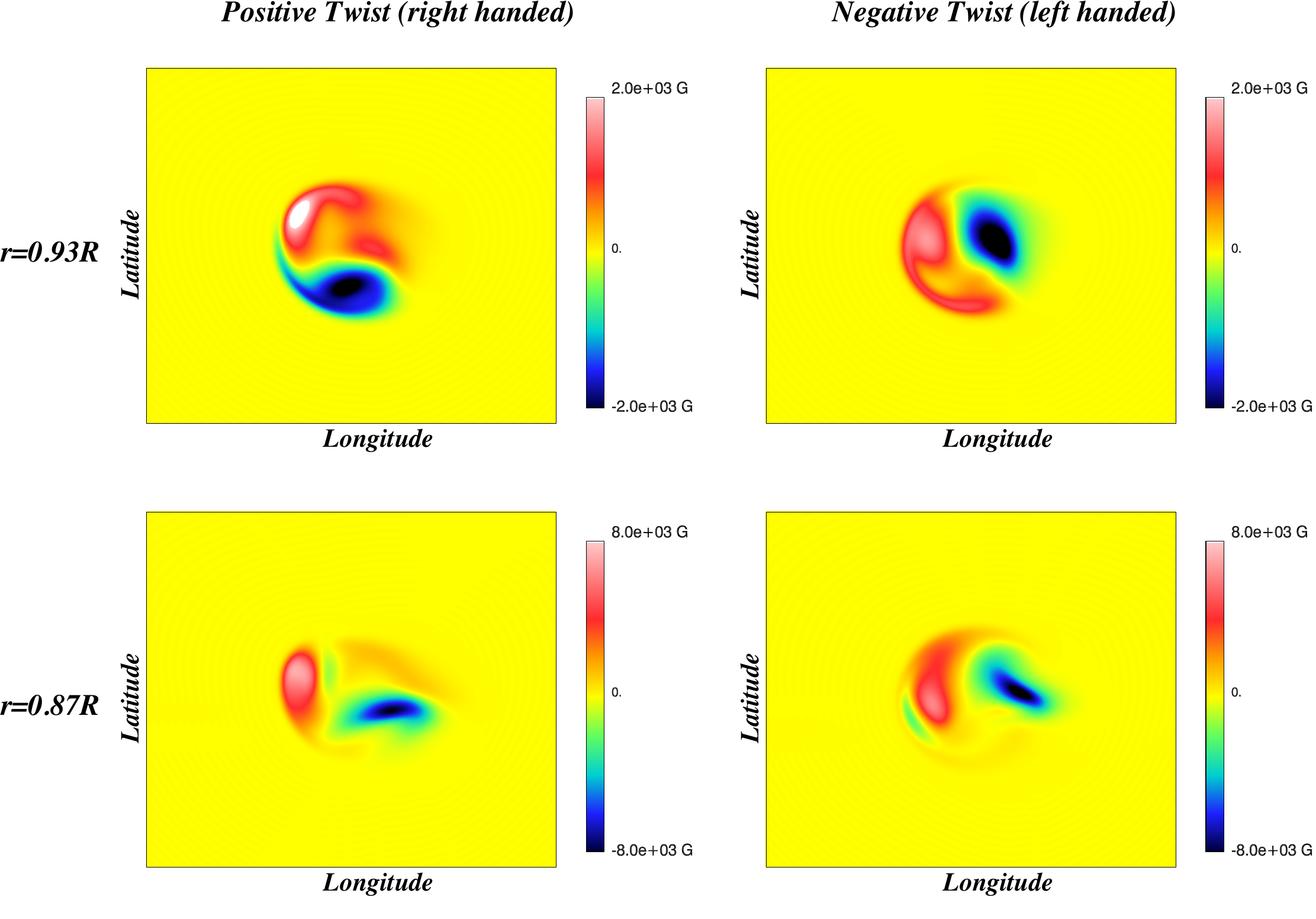}
	\caption{Same as Figure \ref{figure_shsl0} but for Cases 1TwP and 1TwN where $B_0=5 \times 10^4 \rm G$.}
	\label{figure_shsl}
\end{figure*}

 Figures \ref{figure_shsl0} and \ref{figure_shsl} show a zoom on the emerging region in both cases by the time the apex of the loop reaches the top of our domain ($0.96 R\odot$). Blue indicates negative radial field and red positive. The first thing we note is the typical evolution of the inclination of the bipolar structure. Starting from north/south because of the twist of the loop apex, it becomes more and more east/west as we see the legs of the loop emerging. In agreement with previous thin-flux tube calculations \citep[e.g.][]{DSilva93}, we find that a stronger initial field produces a smaller tilt angle since the magnetic tension will tend to oppose the tilting of the magnetic structure. This effect is visible when comparing the left panels of Figures \ref{figure_shsl0} and \ref{figure_shsl} which differ by a factor 2 in initial field strength.

The tilt angle of the bipolar structure and the effect of the initial twist is consistent with previous simulations, especially of \citet{Fan08}. Indeed, in the left-handed case (which is thought to be dominant in the Northern Hemisphere), the tilt angle is not in agreement with observations which tend to show an inclination of the bipolar spots much closer to what we get in the right-handed twist. It was argued previously \citep{Fan08} that to reconcile simulations with observations, a very weak left-handed twist had to be introduced so that the Coriolis force acting on the loop would be strong enough to change the tilt angle from positive to negative sign. The problem is that such a weak twist, not sufficient to maintain a full coherence of the magnetic structure, leads to a strong flux loss and such structures would be likely to have difficulties to rise in a convective environment.

If we look at Figure \ref{figure_shsl} where the initial field strength is decreased to $5\times 10^4 \rm G$ but the amount of twist kept the same, we manage to get a case where the initial left-handed twist is strong enough to maintain the coherence of the loop and where the tilt angle of the emerging radial field gets in better agreement with observations. The Coriolis force has been able to act sufficiently on this loop to rotate the two opposite polarities clockwise and produce a final tilt angle of the same sign in both cases. 

\begin{figure}[h!]
	\centering
	\includegraphics[width=7.5cm]{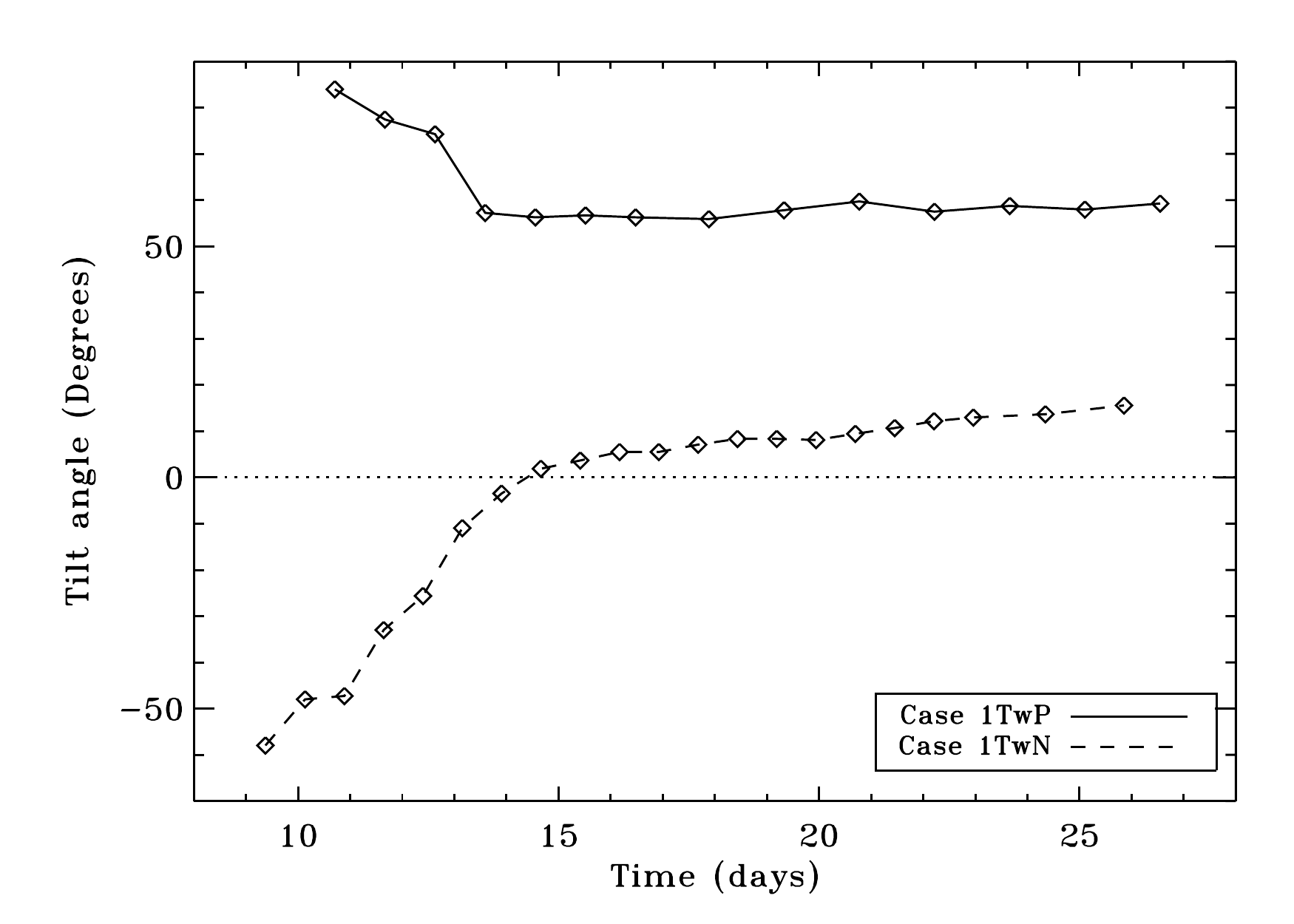}
	\caption{Measure of the tilt angle of emerging bipolar spots for Cases 1TwP and 1TwN, at $r=0.93 R_\odot$.}
	\label{figure_tilt}
\end{figure}

To illustrate this, Figure \ref{figure_tilt} shows on the same plot the temporal evolution of the tilt angle measured in Cases 1TwP and 1TwN, at $r=0.93 R_\odot$. The tilt angle was here determined by locating the peak values of each polarity and measuring the angle between the line linking them and the East-West direction. At the beginning of emergence, the angle is close to $90^o$ for both cases, but with opposite signs. As the emergence progresses, the absolute value of the tilt reduces in both cases. At this stage, the effect of the twist is dominant and corresponds to what is measured here. After about 14 days of evolution, the rapid change of tilt stops and the effect of the Coriolis force starts to be visible, especially on Case 1TwN. As a consequence, the curve in Case 1TwP seems to saturate around the value of $55^o$ and starts to increase again slightly because of the rotation of the spots, to reach a value of about $60^o$ after 26 days of evolution. In Case 1TwN, the tilt angle continues to increase and crosses the $y=0$ line after about 15 days. At this instant, a change of sign has occurred and the tilt has thus become compatible with observations. A final value of about $15^o$ is reached after $26$ days when diffusion then starts to act, which lies in the acceptable range of values when compared to observations at this latitude.

\begin{figure}[h!]
	\centering
	\includegraphics[width=7.5cm]{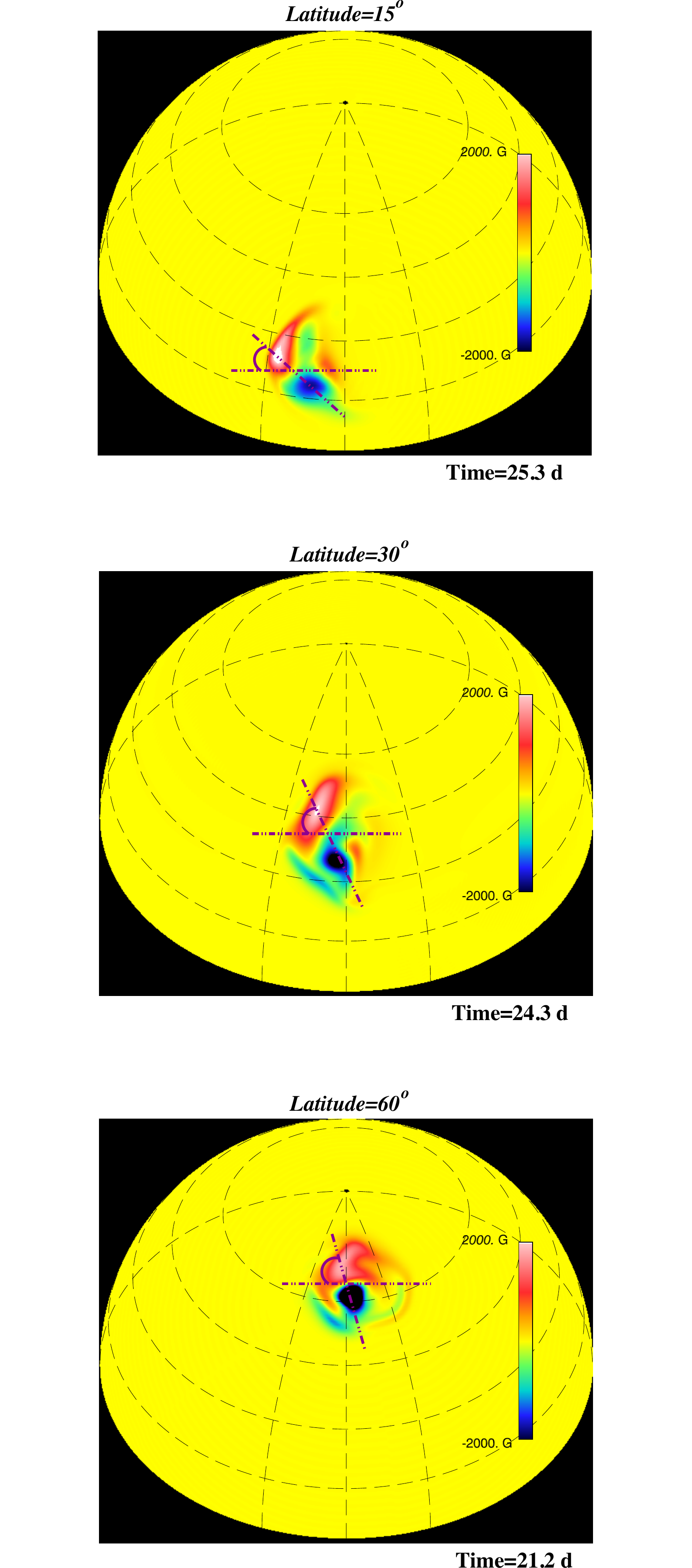}
	\caption{Measure of the tilt angles of cases where the loop was introduced at the latitudes of $15^o$ (Case 1TwP15), $30^o$ (Case 1TwP) and $60^o$ (Case 1TwP60). The Northern Hemisphere is represented centered at $90^o$ in longitude and $45^o$ in latitude.}
	\label{figure_shsl4}
\end{figure}

We now turn to study the influence of latitude on the tilt angle. As we expect and as seen in Figure \ref{figure_shsl4}, the tilt angle depends on the initial latitude since the Coriolis force depends on the latitude. From the estimate of \citet{DSilva93}, the tilt angle is a decreasing function of the cosine of the colatitude (or an increasing function of latitude), as seen in Figure \ref{figure_shsl4} and as seen in observations. As we saw before in the left-handed case, the bipolar spot will have the tendency to rotate clockwise because of the Coriolis force acting differently on the two legs of the loop. As a consequence, the orientation of the bipolar spots in the case where it was introduced at high latitudes ($60^o$) is close to north/south, and not only because we only see the twist of the field lines here but rather because the axis of the loop has been rotated clockwise by about $75^o$. When the initial latitude is $15^o$, the typical tilt angle we get is around $35^o$ while it increases to $60^o$ when a loop at initially $30^o$ is considered. Moreover, we note on this figure that the emergence occurs at slightly different locations in longitude. The center of the bipolar structure being located at around $75^o$ in Case 1TwP15 and $90^o$ in Case 1TwP60. This is due to the asymmetry of the loop which is much more pronounced in the low latitude cases since the action of the Coriolis force on both legs of the loop is stronger then.

We note here that the values for the tilt angle may appear large compared to the classical observations (between $3$ and $10^o$ typically) and to previous thin flux tube calculations \citep{DSilva93, Caligari95} but as shown before (on Figure \ref{figure_tilt}), an initial left-handed twist, associated with a slow rise allowing the Coriolis force to act on the loop, is able to produce tilt angles of about $15^o$ in a case where the initial latitude was $30^o$. This, in turn, is very compatible with observations at the photosphere. It is possible here to make such a comparison if we make the reasonable assumption that the Coriolis force will not have the time to modify the loop orientation and thus the tilt angle during the fast rise through the last few percents of the convection zone.

We shall keep in mind those results for the convective cases where the spherical shell will now be differentially rotating. Since the tilt angle is also a function of the local rotation rate, the tilt angles we will get in the convective cases may well be different.

\subsubsection{Morphology of Emerging Regions: Tongues and Necklaces}

If we concentrate again on Figure \ref{figure_shsl0}, we immediately notice the typical ``tongue''-shape of the emerging region, especially at a deeper radius $r=0.87 R_\odot$. They are indicated by red and blue arrows in the bottom left panel of Figure \ref{figure_shsl0}.  Those elongated structures, also seen in the observations \citep{Lopez00}, in Cartesian simulations \citep{Fan01, Archontis10} and studied in details in \citet{Luoni11}, are mainly due to the twist of the field lines showing in the map of the radial magnetic field. In particular, the orientation of the tongues (or the direction in which they extend) give an indication of the sign of the twist. In the lower left panel of Figure \ref{figure_shsl0}, we thus see that both positive and negative polarities extend on their left side when it is the opposite for the lower right panel where the sign of the twist has changed. Another interesting feature here is the appearance of sharp structures around the emerging region, forming an annular shape that we will from now on call a ``magnetic necklace''. This annular shape of emerging active regions is also often seen in observations but has never been reported as identified magnetic structures, different from the so-called tongues. This is true for observations with experiments and instruments as varied as the Flare Genesis Experiment \citep{Pariat04}, \emph{Hinode} \citep{Otsuji11} or \emph{SOHO}/MDI \citep{Liu06}.

\begin{figure*}[h!]
	\centering
	\includegraphics[width=15cm]{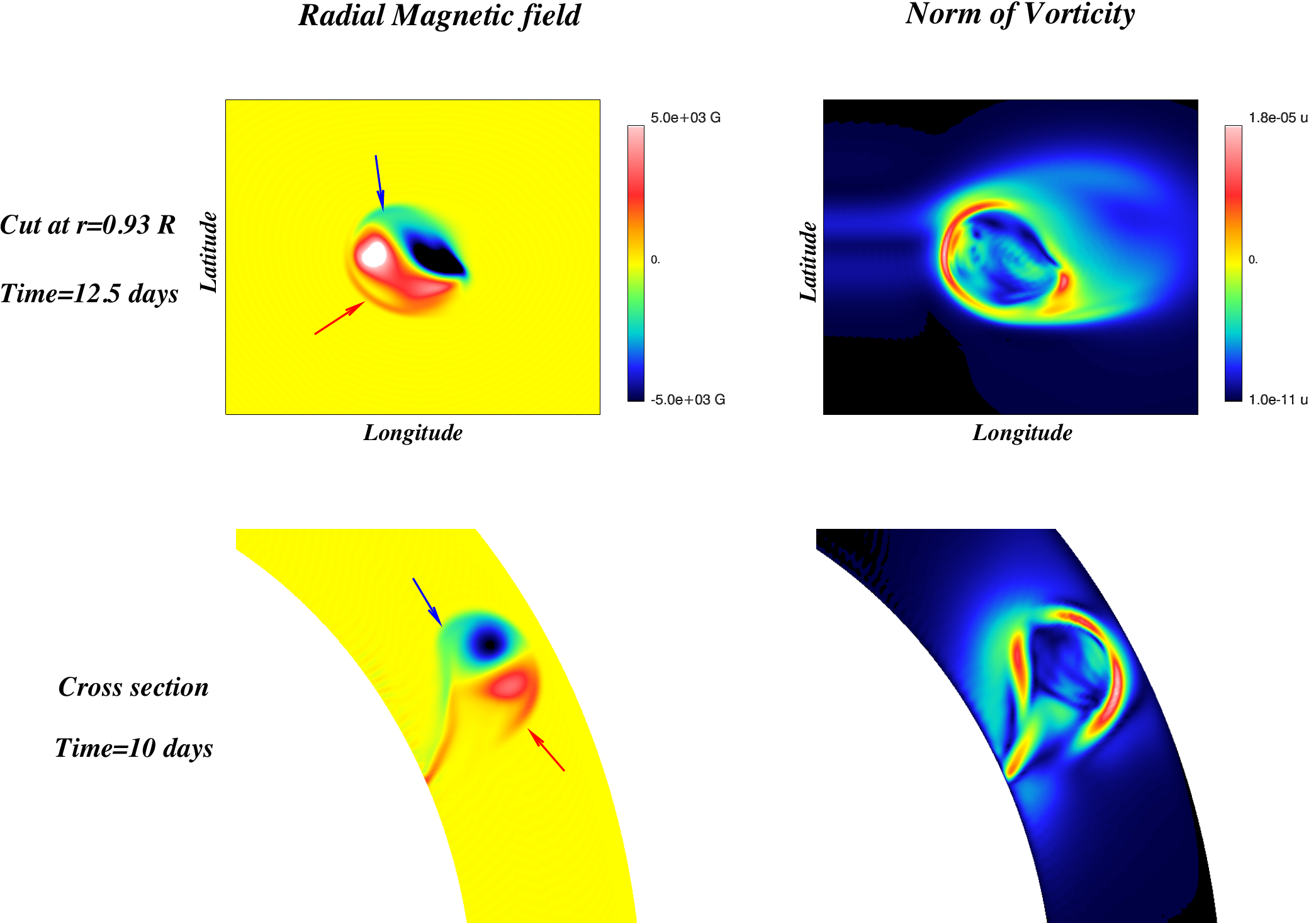}
	\caption{Radial magnetic field and norm of the vorticity in a $(\theta,\phi)$ plane (upper panels) and a $(r,\theta)$ plane (lower panels), focusing on the sharp structures (indicated by arrows) which surround the emerging regions.}
	\label{figure_neck}
\end{figure*}

Those structures are clearly visible in the lower panels of Figure \ref{figure_shsl0} but do not really show in Figure \ref{figure_shsl} because of the low field strength and the rather large magnetic diffusivity in our simulation. Indeed, since those structures are organized at rather small scales, it is necessary to have low enough diffusivities to be able to see those structures at the time of emergence and during their rise. As a consequence, we decided to compute Case 1TwN with a lower magnetic diffusivity (we choose $Pm=5$ here) to see if these small-scale structures could be recovered in this case. Indeed, as shown on Figure \ref{figure_neck}, the magnetic necklace is visible on the zoom in the $(\theta,\phi)$ plane on the emerging region at $0.93 R_\odot$ and is also seen in a meridian cut of the loop during its rise (see red and blue arrows in the left panels). They consist in sharp elongated structures (not to be confused with the tongues) lying at the periphery of each polarity and of the same sign that create a ring inside which the two main regions of strong radial field concentrate. If we compare the regions of appearance of these sharp structures around the bipolar spot with the norm of the vorticity (or enstrophy), it is rather clear that those structures correspond to a velocity shear which produces a strong vorticity. As shown and studied in \citet{Emonet98}, a sharp interface is indeed formed at the boundaries of the flux rope while it rises. This region is the site of vorticity generation via the presence of a magnetic field during the whole loop evolution. Another source of vorticity exists within the tube interior and is linked to the gravitational torque applied between the tube center and its periphery. This process creates the counter-rotating vortices which can completely break up the tube during its rise if the twist (and thus the magnetic tension) is not strong enough, as seen in Section \ref{twist}. Looking at the four panels of Figure \ref{figure_neck}, we understand that the necklace, associated with vorticity generation at the loop edges, will be visible only after a significant part of the loop has emerged. Indeed, since the main vorticity generation is located at the sides of the loop, we need to wait until the axis of the loop crosses the surface $r=0.93R_\odot$ before being able to observe the sharp magnetic structures. By that time, the tongues have thus already significantly faded away and reconcentrated around the two polarities. We can thus expect to see those structures in actual observations only after a certain amount of time after the first signs of emergence. 

\subsubsection{Field Strengths and Fluxes}

We end this section by noting another difference between Cases 1 and 3. In Case 3, the loop seems to maintain more coherence while it rises, both polarities are well-separated, even at the end of the simulation when the bipolar region reaches its maximum extension. This is less the case for the weaker field of Case 1. Indeed, both polarities seem to get mixed and the radial magnetic field lines get more easily advected by the velocity field created by the loop itself (we remind that there is no convection in this case). At the end of the simulation, the magnetic field gets advected at the boundary of the positive radial velocity pattern it creates and finally diffuses away.

\begin{figure*}[h!]
	\centering
	\includegraphics[width=7.5cm]{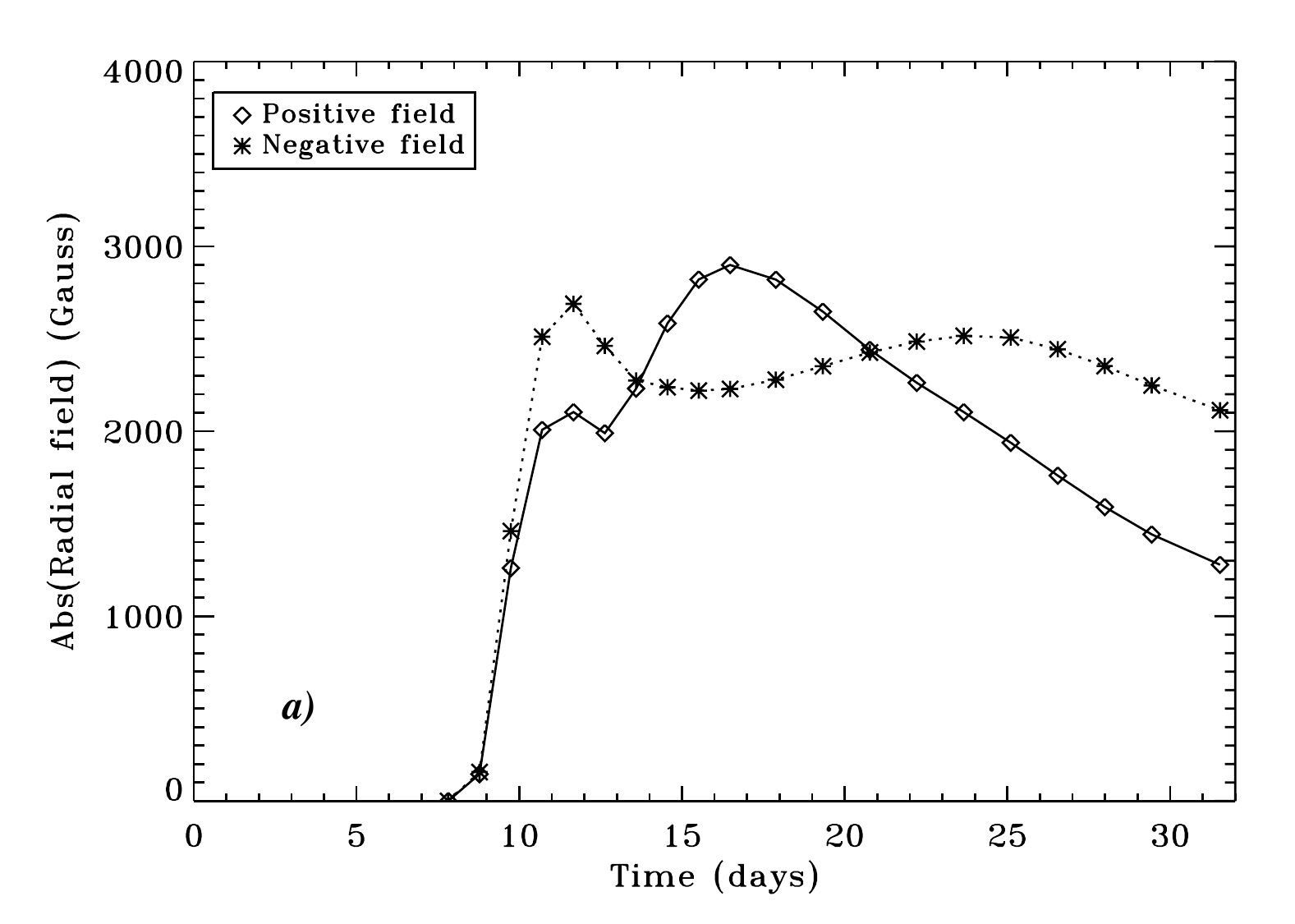}
	\includegraphics[width=7.5cm]{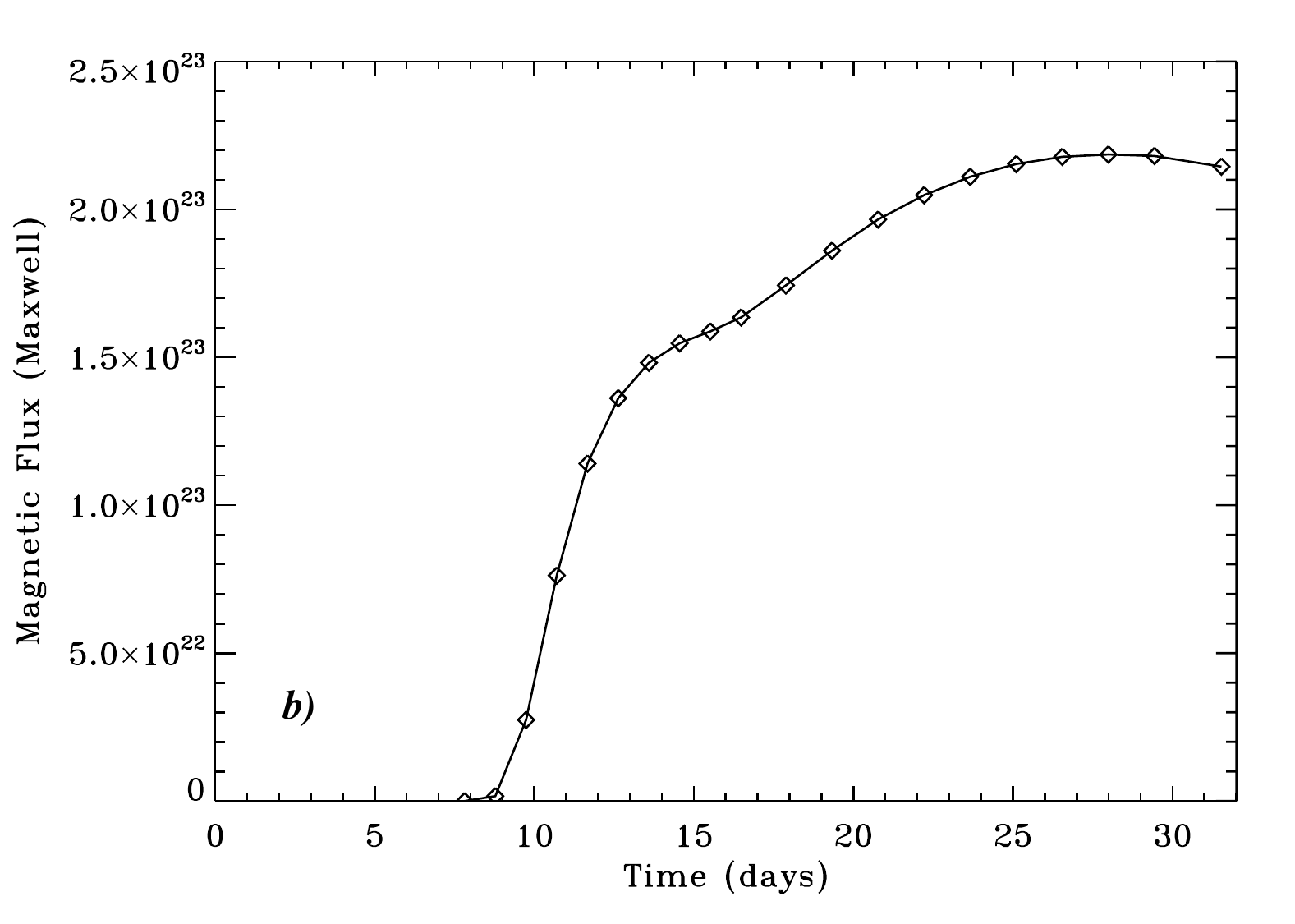}
	\includegraphics[width=7.5cm]{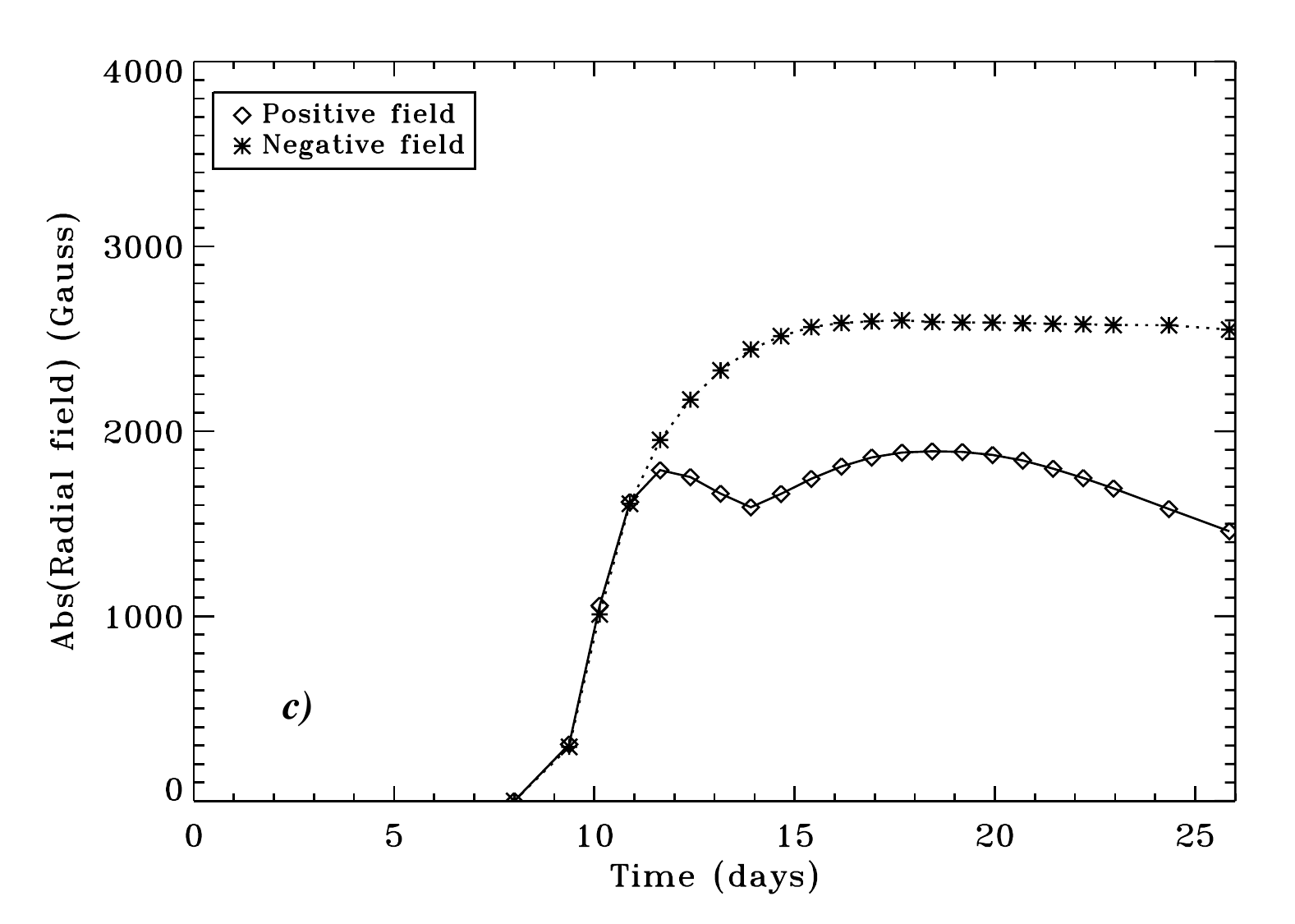}
	\includegraphics[width=7.5cm]{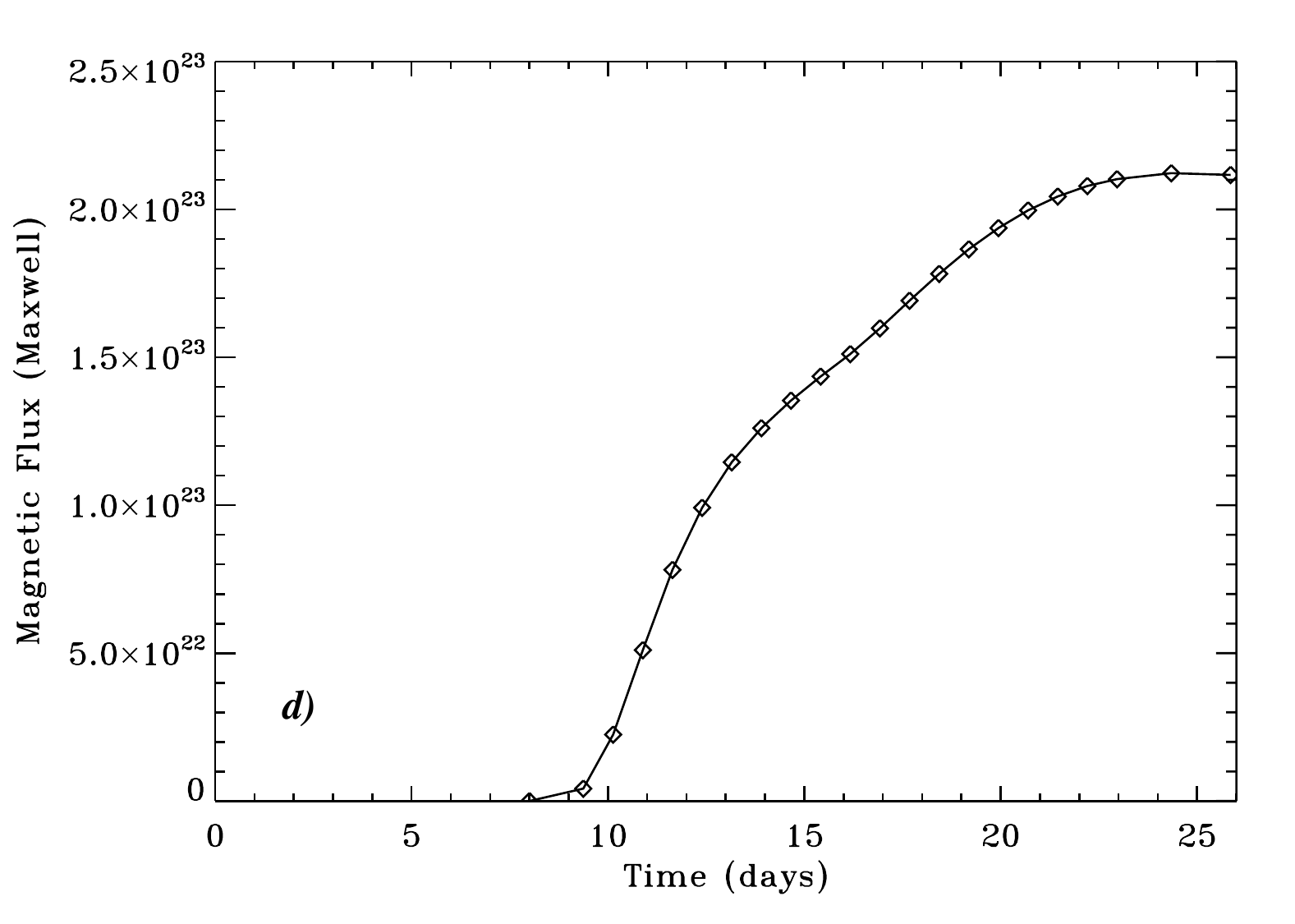}
	\caption{Radial field and unsigned radial flux for Case 1 with right-handed (panels ($a$) and ($b$)) and left-handed (panels ($c$) and ($d$)) initial twist. The values of the field are measured at $r=0.93 R_\odot$ in the center of each polarity where the field is strongest and the fluxes are the integrals of the radial field contained in the trailing (or leading) polarity through a surface enclosing the emerging region.}
	\label{figure_shsl3}
\end{figure*}

If we now focus on this more realistic Case 1 (with a field strength of $5\times 10^4\rm G$, and a typical rise time of more than 10 days), we can move to a more quantitative study of the evolution of the intensity of the radial field emerging in each spot and of the amount of flux it represents. The results of this study are shown in Figure \ref{figure_shsl3}. Panels ($a$) and ($b$) show the evolution, during emergence at $r=0.93R_\odot$ of the radial field and radial flux in Case 1TwP and panels ($c$) and ($d$) in Case 1TwN. In both cases at the beginning of emergence, the negative field in the leading polarity dominates over the trailing polarity (by about $35\%$), while the amount of flux is the same in both polarities (since all emerging field lines of one polarity connect to the other one in the domain of integration). This suggests that the leading polarity is stronger in intensity and more concentrated and the trailing polarity is weaker and less coherent, which is in agreement with observations of emerging active regions \citep[e.g.,][]{Koso08}. Moreover, this feature was also first seen in Cartesian simulations and explained in \citet{Fan93} by the evacuation of plasma out of the leading side of the loop into the following side. The leading side then shrinks and the magnetic pressure increases to balance the same external pressure, resulting in a more intense field in the less extended leading polarity. As \citet{Fan93} points and as discussed also in \citet{Caligari98}, the anchoring of the field lines outside the buoyant part of the loop is a key ingredient to produce a strong asymmetry in field strength. It is also what we find here in Case 1TwN, for example, where a strong asymmetry in inclination develops because of the anchoring of the field lines outside the most buoyant part and where we find a leading leg about 1.4 times more intense than the trailing leg on average (see panel ($c$) of Figure \ref{figure_shsl3}.

We note that the values of the flux are quite strong compared to observations (which report values of about $10^{21}$-$10^{22}$ Mx). This is due to the scales of our magnetic structures which overestimate the size of a flux tube initially located at the base of the convection zone. The initial flux in the magnetic structure in Case 1 is indeed of $6\times 10^{23} \rm Mx$. Simulations with flux tubes of radii about $10$ times smaller, numerically very challenging, would lead to values for the fluxes much closer to observations \citep{Schrijver94}. It is however important to note that despite the discrepancies between the absolute values of the fluxes, due to numerical limitations, the shape of the temporal evolution of the fluxes is in good agreement with observations. Indeed, the flux in each polarity of the bipolar active regions is observed on MDI magnetograms for example \citep{vanDriel03} to increase sharply in the first days of emergence, before saturating and slowly starting its diffusive decay. We get exactly the same kind of evolution in our simulations, visible especially in panel ($b$) of Figure \ref{figure_shsl3}. The flux strongly increases from $0$ to $1.5\times 10^{23} \rm Mx$ in the first 5 days before saturating around this value. In the figure and especially in panel ($d$), we distinguish first a slow rising phase between 8 and 9 days followed by the very sharp increase of the amplitude of the flux after $10$ days of evolution. The first phase is related to the very beginning of the emergence when only the apex of the loop is visible and the only contributor to the flux is the radial magnetic field coming from the twist of the field lines. As the emergence proceeds, however, the feet of the loops where the strong radial magnetic field lies start to emerge and form the bipolar region. It is thus this fast emergence phase of the arched structure which produces the sharp increase of the simulated (and possibly of the observed) magnetic fluxes. As far as the magnetic field strength is concerned, we can wonder what typical horizontal field strength we would get at the photosphere, starting with our flux tubes located at the base of the CZ, in the same way as \citet{Cheung10}. If we follow the evolution of the axial field intensity in time (and thus in radius), we find that the scaling with the reference density is close to $B\propto(\bar{\rho}+\rho)$ (similar to what is found in \citealt{Pinto12}), corresponding to an expansion of the loop in the directions transverse to its axis. At the top of our domain, the typical toroidal (or horizontal) field is of the order $10\rm kG$ (for Case 1TwP for example) and the density $\bar\rho$ of the order $4\times 10^{-3} \rm g.cm^{-3}$. As a consequence, if we consider a density of $4\times 10^{-7} \rm g.cm^{-3}$ at the photosphere, the horizontal field strength would reduce to $1 \, \rm G$, much weaker than the typical observed values (around $100 \rm G$). However, the expansion in the last few percent of the convection zone might be modified (becoming mostly horizontal, for example, due to a decrease of the plasma $\beta$) and then the scaling with density could change drastically and become $B\propto \sqrt{\bar{\rho}}$ as shown in \citet{Cheung10}. We should then keep in mind that we are not modeling the uppermost layers of the convection zone and that it is not relevant here to relate directly the field strengths in our bipolar regions to the field strengths which would be reached at the photosphere. 

Comparing the left-handed and right-handed twist cases, we find that the evolutions, especially for the radial field, are slightly different. Indeed, the dominant leading polarity at the beginning of emergence in Case 1TwP (panel ($a$)) saturates at the value of $2700 \rm G$ and then drops to the value of $2300 \rm G$ while the trailing polarity continues to increase in intensity. At $t=14$ days, the positive trailing polarity thus becomes dominant, before decreasing at $t=21$ days. This is not the case for Case 1TwN (panel ($c$)), where the leading polarity always stay dominant in strength and approximately equal to the trailing polarity in flux. This could be due to the asymmetry built up during rise, which is smaller in the left-handed twist case. This is indeed seen on cuts at constant latitude as in Figure \ref{figure_eqsl} (not shown here). As a consequence, in Case 1TwP, the asymmetry between the legs is strong, the leading polarity reaches its maximum strength and thus starts to diffuse away before the trailing polarity. This explains the decrease in amplitude in Figure \ref{figure_shsl3} first seen for the negative field and then for the positive field. In Case 1TwN, the peak positive and negative radial fields are measured at $r=0.93 R_\odot$ at approximately the same time since the asymmetry is less pronounced and they both saturate at the same time (at the value of $1900 \rm G$ for the positive polarity and $2500 \rm G$ for the negative one), before starting their diffusion.

\section{EVOLUTION OF A SINGLE LOOP WITH CONVECTION}
\label{sect_conv}

\begin{figure*}[h!]
	\centering
	\includegraphics[height=12cm]{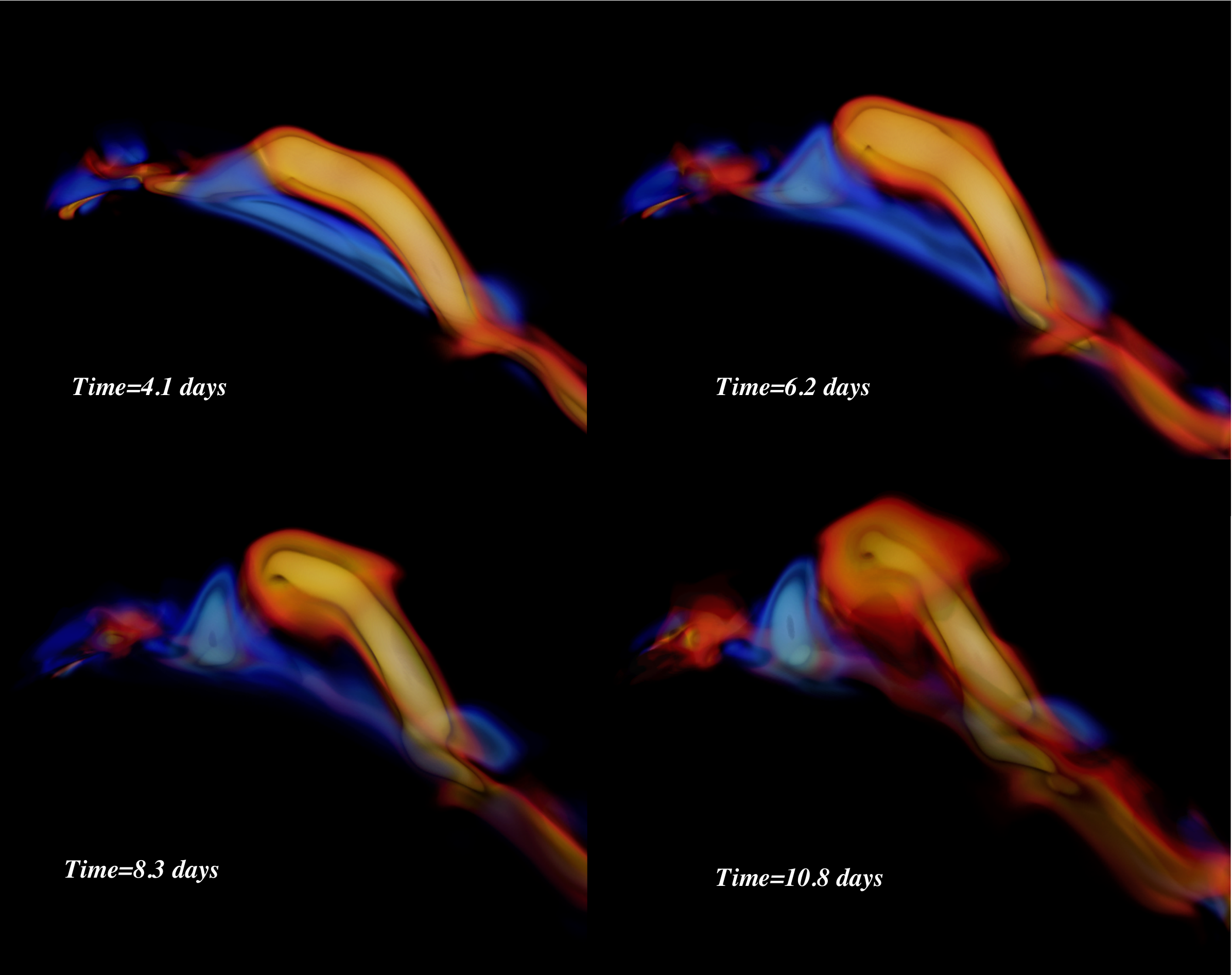}
	\caption{Volume rendering of $B_{\phi}$ while the loop rises through a convective layer. As in Figure \ref{figure_isen3D}, the loop is viewed from below, looking up at the North Pole.The deformation of the loop by convective motions is clearly visible here when compared to Figure \ref{figure_isen3D}.}
	\label{figure_conv3D}
\end{figure*}

We now turn to investigate the behavior of a loop similar to the cases presented above, but inside a convective background. As in the global flux tubes simulations \citep{Jouve09}, the buoyancy force will compete with the downflows to drive the loop towards the top of the computational domain and the emerging region will have significantly different characteristics than in the isentropic cases. Figure \ref{figure_conv3D} shows the typical three-dimensional evolution of an $\Omega$-loop evolving in a convective environment, this corresponds to Case 2TwP. We only see here the evolution of the toroidal field but a turbulent convective velocity field exists all around it. Moreover, differential rotation and meridional circulation have well-established profiles here, as stated in Section \ref{sect_hydro}.

We recover some aspects of the cases studied in Section \ref{sect_isen}, like the asymmetry between the legs and the creation of a negative toroidal field at various regions at the loop periphery. Again, a local $\Omega$-effect is at work here, with a slightly more complex structure since the velocity field is now organized at different spatial and temporal scales. But the main ingredients (azimuthal velocity gradients and radial and latitudinal components of the magnetic field) are still the same and able to locally create negative toroidal field.

\subsection{In the Bulk of the Convection Zone}
\label{sect_bulk}

\begin{figure*}[h!]
	\centering
	\includegraphics[height=11cm]{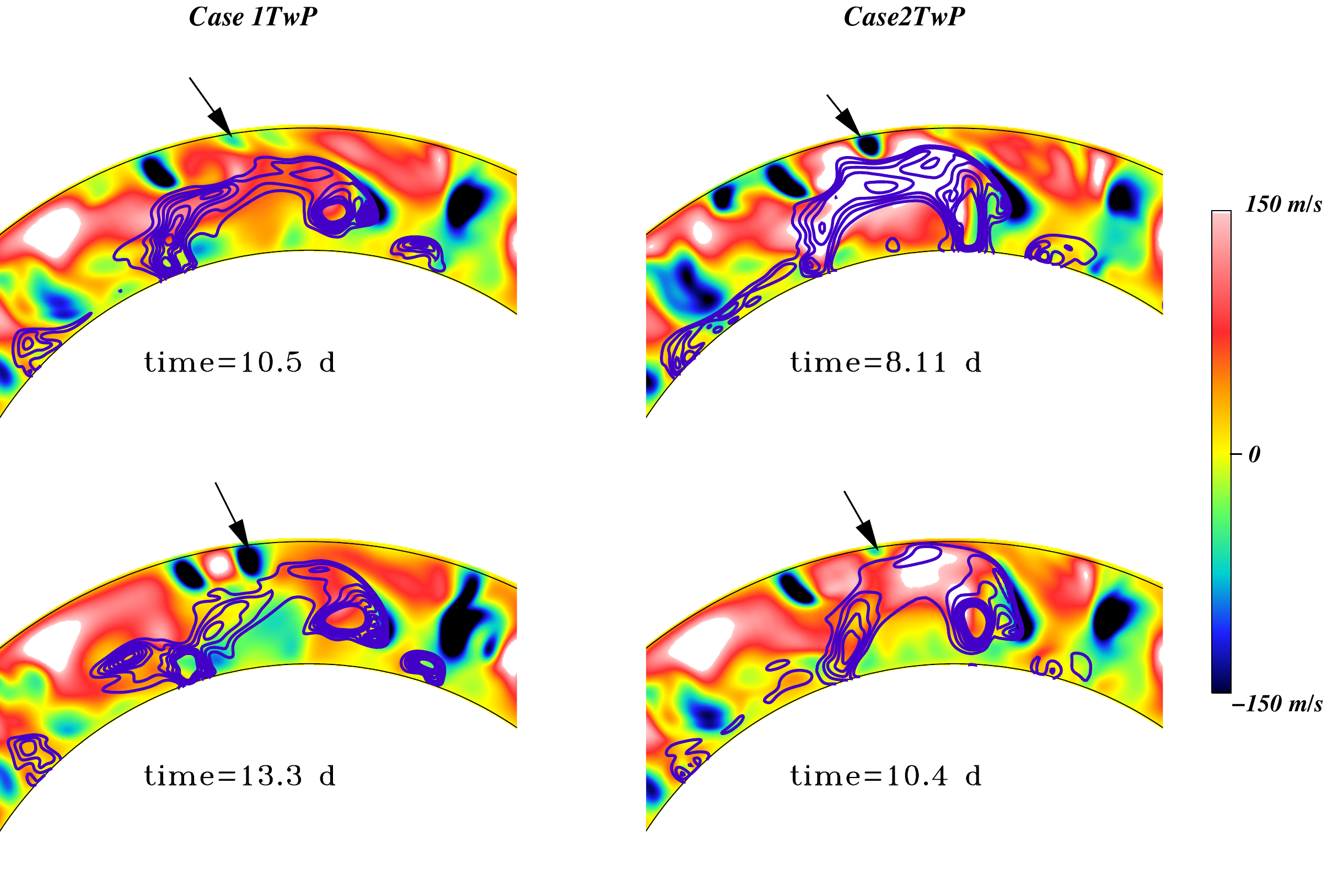}
	\caption{Cut at the latitude of $30^o$ of the magnetic energy (contours) superimposed to the background radial velocity (blue meaning downflows and red meaning upflows) for Cases 1TwP and 2TwP. The color bar indicates the values taken by the radial velocity.}
	\label{figure_eqslc}
\end{figure*}

In such fully convective setting, we expect the initial field strength in the flux tube to be an important parameter for the evolution and rise of the magnetic field. Indeed, as shown in previous work involving a uniformly buoyant flux tube \citep{Fan03, Jouve09}, the downflows and upflows of the convective layer control the rise velocity of the tube and are sometimes even able to pin it within the convection zone, resulting in structures emerging at specific longitudes. Here again, we get the same kind of competition between convection and magnetic buoyancy, which can be seen on Figure \ref{figure_eqslc}. This figure shows the evolution of the loops of Case 1TwP and Case 2TwP. They correspond to the second and third panels of Figure \ref{figure_eqsl}, for which the asymmetry was clearly modified when the parameters were changed such as to have more intense field. In Case 1, the initial field strength is $5\times 10^4\rm G$, i.e., slightly below the equipartition field strength of $6\times 10^4\rm G$, corresponding to the magnetic field in equipartition with the strongest downdrafts at the base of the CZ. In Case 2, the initial magnetic field strength is $10^5\rm G$, corresponding to $1.7$ times the equipartition field. We superimpose here the magnetic energy contours to the background radial velocity (representing the convective motions). For Figure \ref{figure_eqsl}, we chose to represent the loops at about $t=10$ days since their position in the bulk of the CZ was not very different for the isentropic cases. Here, at about $t=10$ days, the loops have not reached the same height, the apex of the loop of Case 2TwP has reached $r=0.93 R_\odot$ after about $8$ days already, whereas the loop of Case 1TwP reaches the same height at $t=13$ days and was thus significantly slowed down compared to the isentropic case. This can be explained by the structure of the flow in which the loops are embedded. In Case 2TwP and as seen especially on the panel at time $t=8.11$ days, the velocity field associated to the Lorentz force is dominant. At the location of maximum magnetic field, a strong upflow thus develops (a maximum positive radial velocity of $350 \rm m.s^{-1}$ is reached here), the loop is dragged upward and the influence of the downflow shown by the black arrow is limited. The rise of the loop is thus widely promoted by the strong upward convective cell produced by the intense magnetic field. On the contrary, in Case 1TwP where the magnetic energy content (and thus the buoyancy) is weaker, the radial velocity created by the presence of a magnetic field is of the same order as the background convective motions and the downflows (as the one indicated by the arrow) are much more efficient at maintaining the magnetic loop under the surface and preventing the axis from emerging.  

Another difference we note here compared to the isentropic cases is the degree of asymmetry between the trailing and leading legs. The strong asymmetry visible in Case 1TwP when the loop was embedded in an isentropic background is still present at $t=13.3$ days but less obvious since the advection by the convective motions has also mixed the field lines and significantly modified the structure. However, in Case 2TwP at $t=10.4$ days, the same degree of asymmetry as in the isentropic case is found, showing again that this regime is dominated by the Lorentz force, contrary to Case 1TwP where the magnetic structure is essentially frozen in (advected by) the convective motions. 

\begin{figure}[h!]
	\centering
	\includegraphics[width=7.5cm]{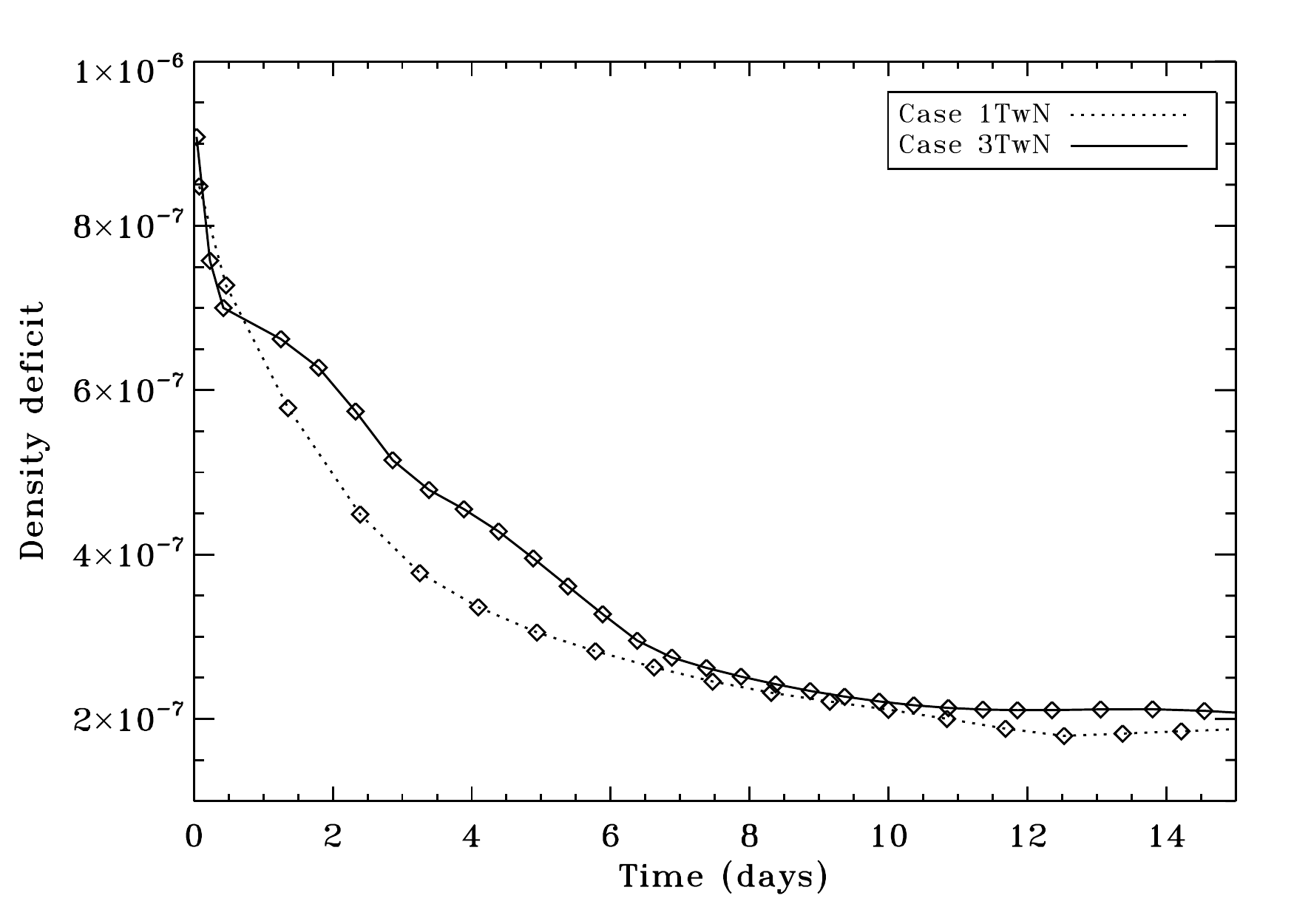}
	\caption{Density deficit at the apex of the loop during its rise in Cases 1TwN (dotted line) and 3TwN (solid line). The internal density is measured inside the tube apex, where the magnetic field is maximum and compared to the external density which is taken to be the average value of the density at the corresponding depth.}
	\label{figure_density}
\end{figure}

In order to gain a better insight on why we get such a different behavior between what we will now call the \emph{convection dominated} regime of Case 1TwP and the \emph{magnetic field dominated} regime of the other cases, we plot the evolution  of the density deficit inside the loop at its apex as a function of time in Cases 1 and 3. The results are shown in Figure \ref{figure_density}. They correspond to Cases 1TwN and 3TwN but the results are similar for the right-handed twist cases. Our results here and especially the distinction between convection and magnetic field dominated cases are similar to what is obtained in the recent thin-flux tube simulations of \citet{Weber11}. Indeed, they find that an initial field strength above $10^5 \rm G$ ensures an evolution dominated by magnetic buoyancy while a loop initialized with $B_0=1.5 \times 10^4 \rm G$ is strongly influenced by convective motions. Similar values are found here.

Figure \ref{figure_density} shows the evolution of the density deficit inside the loop compared to its surroundings, for Case 1 (dotted line) and Case 3 (plain line). This density deficit represents the buoyancy of the structure and thus its capacity to reach the top of our computational domain. Both loops started with approximately the same density deficit, as seen at the beginning of the evolution. The two cases first follow the same amount of buoyancy loss during about half a day and then the two cases start to have different behaviors. Indeed, in Case 1, the loops continues to smoothly lose its density deficit while the simulation proceeds and follows the exponential decay of a typical solution of diffusion equation. After $t=10$ days, the density fluctuations in the bulk of the convection zone are of the same order as the density deficit in the loop and the magnetic structure stops being buoyant. We also note that convective motions act on the magnetic structure to create smaller scale regions which will thus diffuse faster and lead to a smaller field strength and thus smaller buoyancy. At $t=15$ days, the loop has still not reached the top of the domain and the only reason why we will get some emerging magnetic field in this case is that upflows will drag the loop upward. The coherent bipolar structure will then be mostly lost when the magnetic structures will emerge at $r=0.93 R_\odot$, as we will see in the following section. In Case 3 on the contrary, the evolution of the density deficit is quite different. The major trend is of course a decrease of the buoyancy of the structure but this decrease does not follow a decaying exponential as we had in the previous case. Here, strong currents generated by the higher field strength (initially $10^5\rm G$) will play a role in the energy balance of the loop and the diffusion of entropy will be less efficient in this case. As we see on the figure, the buoyancy of the loop of Case 3 always stay higher than in Case 1. As a consequence, at $t=15$ days, a significant part of the loop has emerged at the surface as a coherent bipolar structure.  The loop thus keeps both its coherence and its buoyancy for a longer time and the structure of the emerging regions will tend to be less influenced by the convective motions.

\subsection{Structure of the Emerging Regions}

As stated in the introduction, one of the main questions we wish to answer in this work is the following: ``what is the influence of convective motions and large-scale flows on the structure of our emerging magnetic field?''. Although our top boundary condition is still at $28 \rm Mm$ below the surface and that density and pressure drops may well cause large modifications of the observed magnetic field, we can get some insight on what type of active regions may be created at the solar surface from our numerical simulations.

\subsubsection{Convection versus Magnetic Field Dominated Regimes}

\begin{figure*}[h]
	\centering
	\includegraphics[width=15cm]{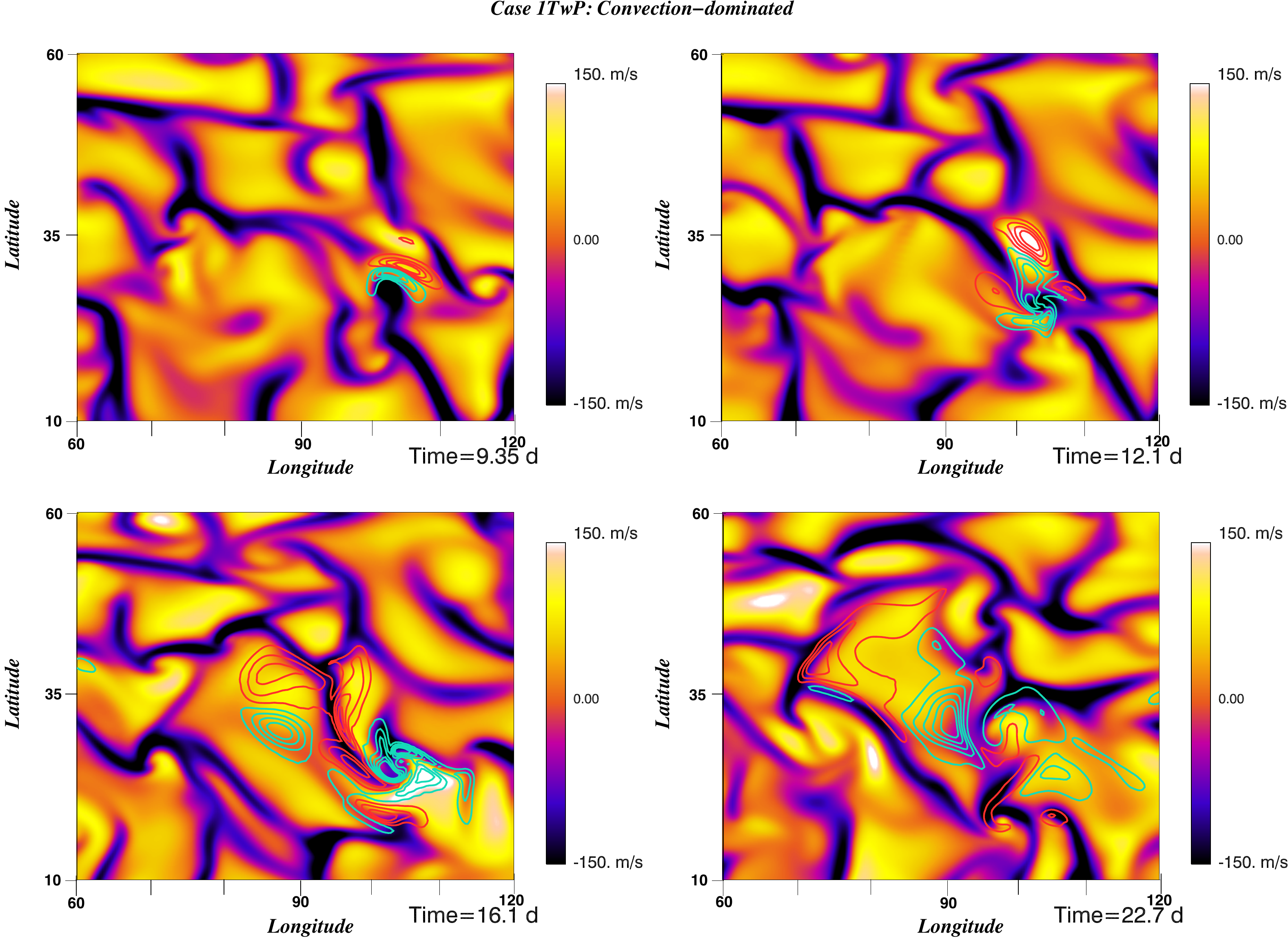}
	\caption{Zoom at $r=0.93R_{\odot}$ on the emerging radial field (red contours represent positive $B_r$) inside the convective environment (yellow colours represent upflows), in Case 1TwP. The typical magnetic field strength in these snapshots is about $2000 \rm G$. }
	\label{figure_shsl1}
\end{figure*}

In this section, we focus on the \emph{convection dominated regime} of Case 1TwP (and 1TwN), where the emerging magnetic field will be strongly affected by the convective motions. Figure \ref{figure_shsl1} shows four snapshots of the emerging region zoomed in latitude and longitude, at $r=0.93R_\odot$. The radial velocity (colors) represents the convective motions (blue downdrafts and yellow upflows) and we superimpose the emerging radial field (red contours indicating positive polarity). At the beginning of emergence at about $t=9$ days, a well-defined bipolar structure can be identified, at the latitude of $30^o$, indicating that the rise was indeed radial and not parallel to the rotation axis. We note already on this snapshot that the strong downflow located at the latitude of $30^o$ and the longitude of $105^o$ strongly interacts with the negative polarity and is responsible for its cashew-nut-like shape. As the emergence proceeds, the well-identified bipolar structure is replaced by a much more complex field structure. The negative polarity gets squeezed into the previous strong downflow, while regions of positive polarity start to emerge all around. The concentration of magnetic field in the downflow lanes is even more visible on the third snapshot at $t=16.1$ days where the field concentration seen on the previous panel continues to accumulate in the downdraft while a new emerging bipolar structure appears at the longitude of $90^o$ where a large upflow was already present and responsible for this new emergence. We then note that this new region gets quickly advected toward the boundary of this large upflow to, in turn, get concentrated into the neighboring downdrafts, leading to an evolution very close to what is observed in simulations of magnetoconvection \citep{Weiss96}. In this case, convective motions were thus very weakly affected by the emerging magnetic field and completely dominates its evolution.

\begin{figure*}[h]
	\centering
	\includegraphics[width=15cm]{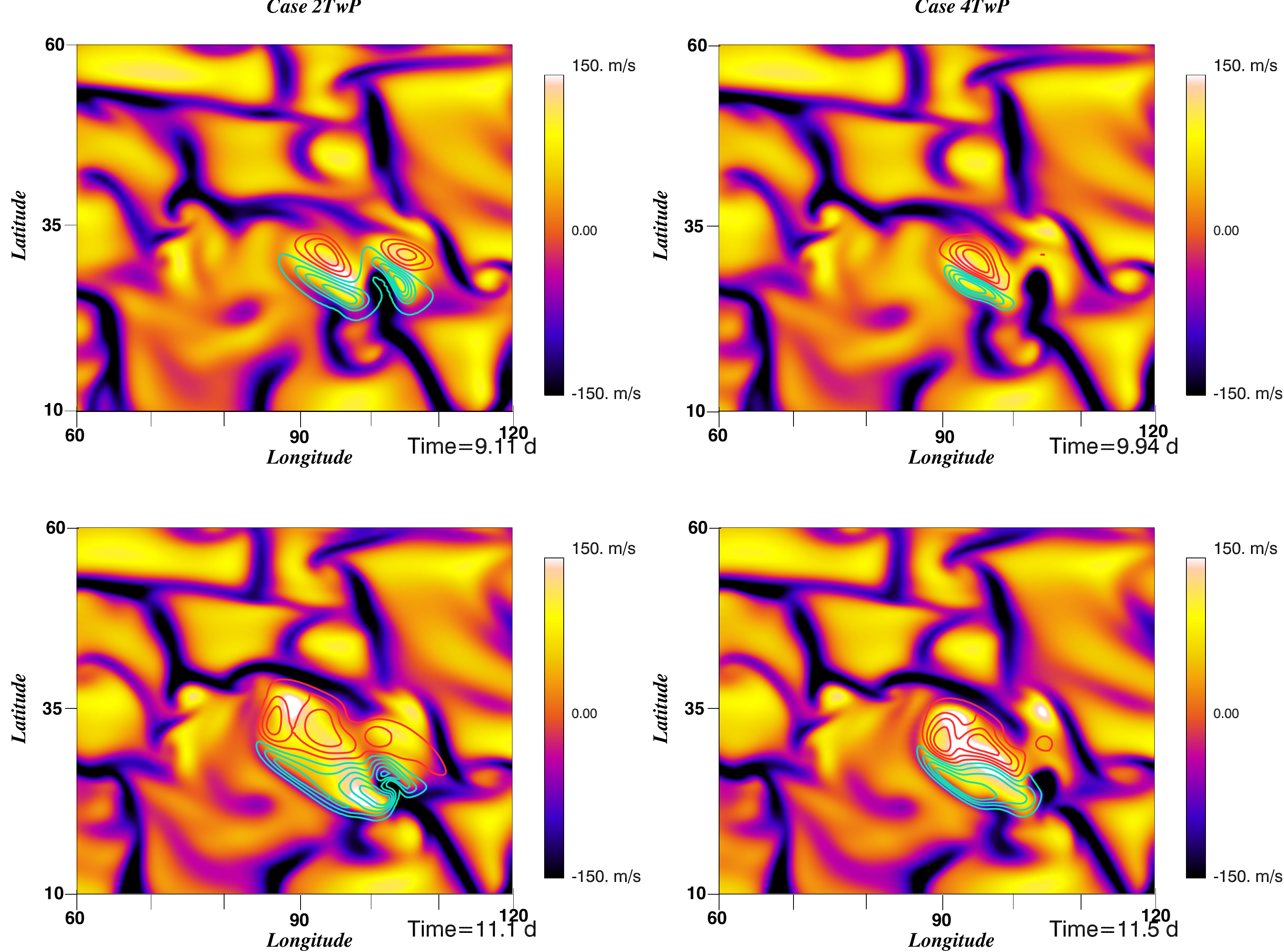}
	\caption{Same as Figure \ref{figure_shsl1} but for Cases 2TwP and 4TwP. The typical magnetic field strengths here are between $5000\rm G$ and $8000\rm G$.}
	\label{figure_shsl2}
\end{figure*}

The evolution shown in the previous figure is quite different when cases with stronger buoyancy and a stronger initial field strength are considered. We now turn to investigate the structure of the magnetic field emerging in Cases 2TwP and 4TwP which initially possess the same buoyancy but different field strengths ($10^5\rm G$ for Case 2TwP and $1.35\times10^5\rm G$ for Case 4TwP). These values respectively correspond to $1.7$ and $2.3$ times the equipartition field strength. The results are shown in Figure \ref{figure_shsl2}. Four snapshots are shown, two for each case, at approximately the same times, $t=9$ days and $t=11$days. Case 2TwP is interesting since it is an intermediate case between \emph{convection dominated} and \emph{magnetic field dominated} evolution. Indeed, at the beginning of emergence, a bipolar structure emerges at about $95^o$ as expected (and as found in the isentropic cases) but another region, which is in fact another part of the loop, also emerges at the longitude of $105^o$. This is due to the strong downflow lane at about $100^o$ of longitude and its neighboring upflow which will have the effect of dragging the loop upwards in the upflow and pinning it down in the downflow (those flows are also visible in Figure \ref{figure_eqslc}). As a consequence, two separate regions seem to emerge when they are part of the same $\Omega$-loop but strongly influenced by convection. As the simulation evolves and emergence proceeds, we recover that the magnetic field gets advected in the downflow lanes at the largest longitudes but shapes as a typical active region at around $90^o$. Indeed, the tongue-shape appears here, although modified by the convective motions. This typical case, intermediate between a \emph{convection dominated} and \emph{magnetic field dominated} evolution, is also interesting since it directly relates to observations of emerging regions. Indeed, the regions of strong magnetic field, in the center of the active regions, are only marginally affected by the convective motions, as opposed to the surrounding plage regions whose structure is affected by the supergranular scales of convection. We could thus consider that the parameters of Case 2 enable us to simulate an emerging region with characteristics typical of an observed active region, in particular its field strength in each polarity, its competition with convective motions, its tongues or its tilt angle.

If we now look at Case 4TwP where the initial magnetic field was stronger, the emergence is much more similar to what we had in the isentropic cases. The orientation of the unique (unlike Case 2) structure is first North-South to become more and more East-West, the tongues are easily identified, a strong upflow is created by the loop itself and dominates its evolution during the emergence process. We observe in Figure \ref{figure_shsl2} that the apex of the loop emerges at the longitude of $95^o$ and that no radial field is visible at $105^o$ in Case 4, contrary to Case 2. This is due to the fact that in Case 2, the advection of the magnetic structure by the convective motions significantly modifies the shape of the loop and thus the asymmetry between the trailing and leading legs. In Case 4, the evolution is again closer to the isentropic case and convection has little influence on the loop and is thus not strong enough to produce an emerging region within the upflow located at $105^o$. To conclude on these cases, although Case 2 is more influenced by convection, a clear impact of the emerging structure on the velocity field is visible in both cases, especially at $t=11$ days where a strong upflow is created at the very location of emergence.

\subsubsection{Effect of Twist and Latitude}

In this last section, we investigate the effect of changing the sign of the twist of the initial magnetic loop, as well as modifying its latitude of introduction. We saw in the isentropic cases that the sign of the twist would influence the tilt angle of the emerging region but that a tilt compatible with observations could still be found with an initial left-handed twist. We now address the question of the possibility to get tilt angles even more compatible with observations when the magnetic loop is embedded in a convective environment.

\begin{figure*}[h]
	\centering
	\includegraphics[width=15cm]{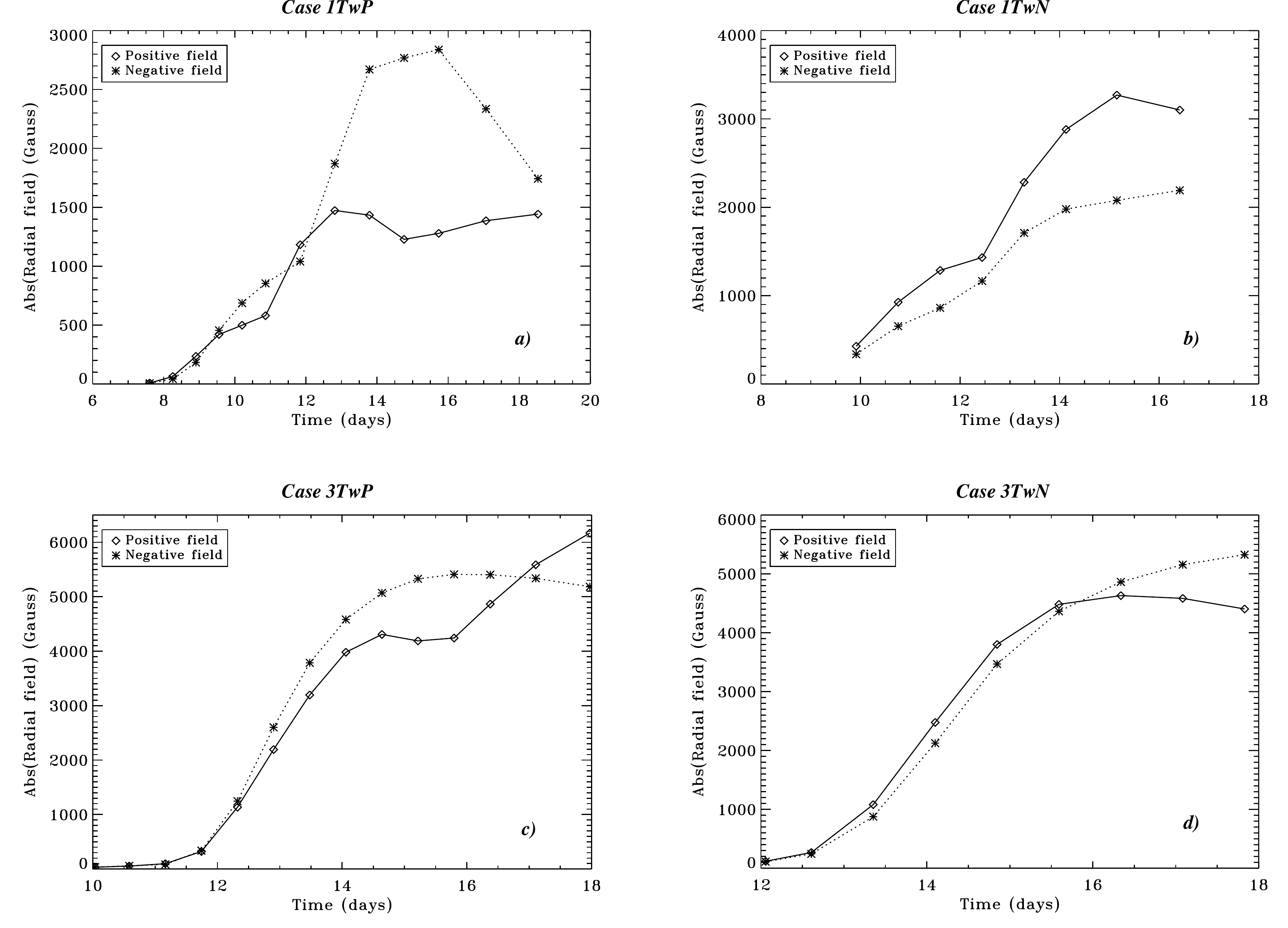}
	\caption{Temporal evolution of the radial magnetic field intensity in the trailing and leading polarities at and right after emergence, for cases 1TwP (panel ($a$)), 1TwN (panel ($b$)), 3TwP (panel ($c$)) and 3TwN (panel ($d$)).}
	\label{figure_br}
\end{figure*}

Before specifically studying the tilt induced by the twisted structures, we investigate the impact of a negative initial twist on the intensity of the emerging radial field. Indeed, it was already found in the isentropic cases that the strength of the emerging polarities and the positive and negative fluxes had a different evolution when a left-handed or right-handed twist was considered. It is still the case when the loops are introduced in a convective shell. Figure \ref{figure_br} presents the evolution of the radial field of the leading and trailing polarities at the time of and after emergence for Cases 1TwP, 1TwN, 3TwP and 3TwN. Similar pictures for Cases 1TwP and 1TwN were shown in Figure \ref{figure_shsl3} for the isentropic simulations. We can thus first compare the convective and isentropic situations for Case 1. The first thing we note and which is true both for the right-handed and left-handed twist, is that the loop in the isentropic case had a quicker rise than in the convective case. Similar results were found for uniformly buoyant flux tubes in \citet{Jouve09}. Indeed, for example at $t=10$ days for the isentropic cases, the typical field strength already reached $2000\rm G$ whereas it barely gets to $500\rm G$ in the convective cases. This is related to the explanations of Section \ref{sect_bulk} where the buoyancy loss was shown and thought to be partially due to the convective motions creating smaller scales and thus stronger dissipation. 

Another striking feature is the strong increase of negative $B_r$ in Case 1TwP (panel ($a$)) at $t=13$ days and of positive $B_r$ in Case 1TwN (panel ($b$)) at about the same time. This is mainly due to the fact that these cases are dominated by convective motions. Indeed, as we saw in Figure\ref{figure_shsl1}, the negative polarity in Case 1TwP between $t=12$ and $t=16$ days gets strongly squeezed by the convective motions in the narrow downflow lanes. This compression of magnetic field tends to produce stronger concentration of fields, explaining the strong increase of negative $B_r$ in Case 1TwP and then its sharp decrease due to fast dissipation of small scales. A similar situation happens in Case 1TwN for the opposite dominating polarity. In Cases 3TwP and 3TwN (panels ($c$) and ($d$)) where the initial magnetic field is twice as large, such sharp increase and decrease of the two polarities are not visible since the loop emerges as a global magnetic structure and is much less influenced by convective motions. It is worth noticing that we may have a possible way here to explain the formation of complex active regions from where powerful (X-class) flares could originate. Indeed, active regions with a complex topology could be created by the turbulent convective motions acting on moderately strong flux tubes (of approximately $5\times 10^4 \rm G$ at the base of the CZ). By the time they reach the top of the convection zone, the flux tubes have been strongly deformed and the subsequent emerging radial field consists in several patches of opposite polarities, as seen in Figure \ref{figure_shsl1}. We could argue that this complex active region, whose progenitor is a relatively weak flux tube, would not be able to produce very energetic flares. However, when panel ($a$) of Figure \ref{figure_br} is considered, we conclude that the converging motions at the sub-photospheric level are able to re-concentrate the magnetic structure in the strong downdrafts, in order to locally produce a strong increase of the radial field. In particular, we note that the maximum positive radial field is about twice as large as the negative $B_r$, which was not the case in the isentropic case where the re-concentration by convective motions could not occur. This scenario may thus explain the injection of energy in the corona by convective motions acting on complex active regions originating from initially weak flux tubes.

As far as the tilt angle is concerned in these simulations, the influence of the initial twist, which is not included in the thin-flux tube calculations, is again crucial. If we follow the position in latitude of the positive trailing and negative leading polarities (not shown here), we find that, at the beginning of emergence, the leading polarity is located at higher latitude than the leading one, unlike in solar observations. However, in Case 1TwN, the clockwise rotation due to the Coriolis force acting on the legs of the loop tends to reverse this situation about $5$ days after emergence, implying a leading polarity at slightly lower latitude than the trailing one and thus to a tilt of about $5^o$, in agreement with Joy's law. This is similar to what was found in the isentropic Case 1TwN. Case 3TwN does not follow the same evolution. Indeed, the orientation at emergence is kept even after $10$ days after emergence before the magnetic field is swept towards the downflow lanes and dissipates. Again, we need a loop that is sufficiently influenced by the Coriolis force to get a tilt in agreement with observations with an initial left-handed twist believed to dominate in the Northern Hemisphere.

\begin{figure}[h]
	\centering
	\includegraphics[width=7.5cm]{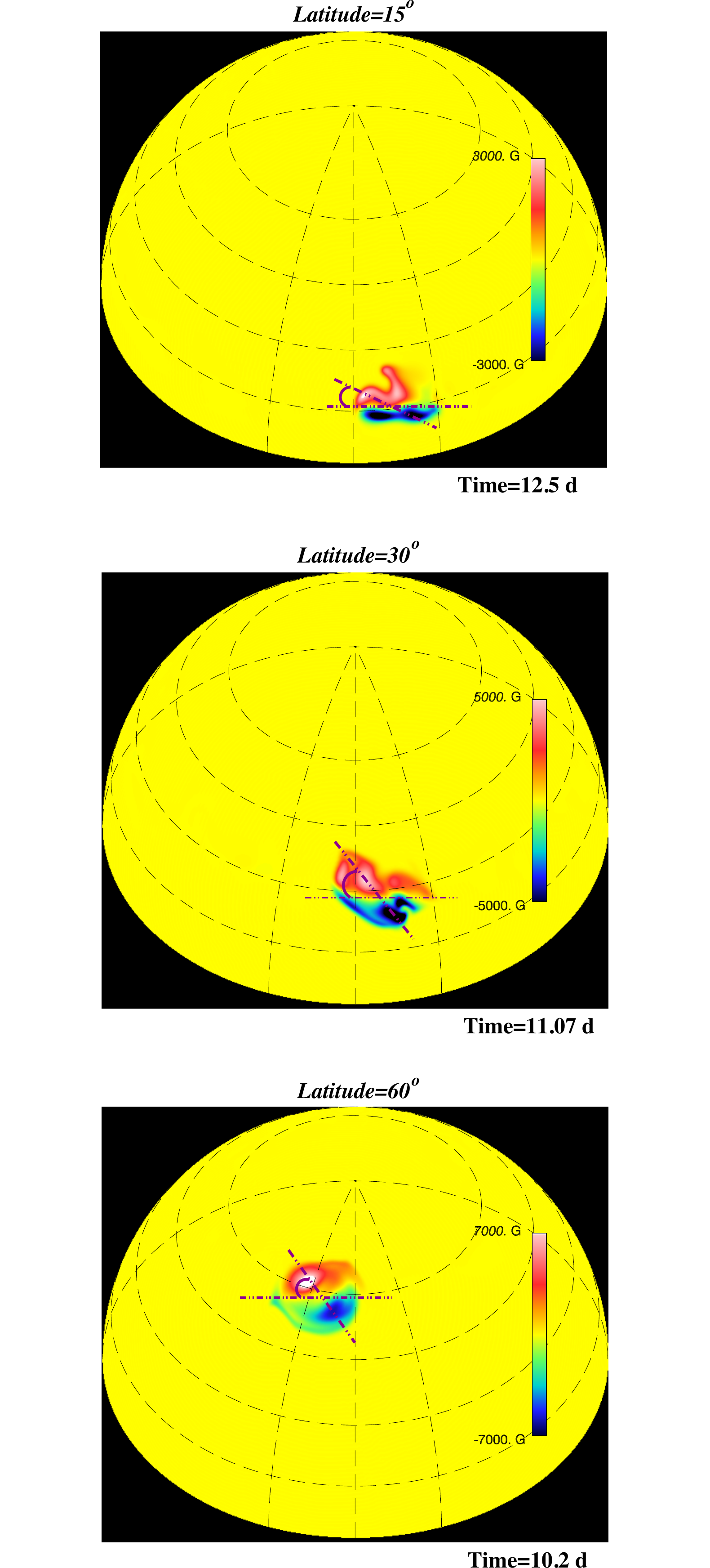}
	\caption{Same as Figure \ref{figure_shsl4} but for the convective Cases 2TwP15, 2TwP, and 2TwP60.}
	\label{figure_convlat}
\end{figure}

Of course, not only will the initial twist play a role in the tilt of the emerging region but also the latitude of introduction and emergence, as already pointed out in the isentropic cases. We thus computed extra cases as in the isentropic simulations, Cases 2TwP15 and 2TwP60. We did not choose the same initial parameters as in the isentropic cases to study the effect of latitude, since Case 1 was computed in the isentropic case and it is believed here to be influenced too much by convection to produce a significant trend and thus to allow drawing any conclusions on the tilt at the time of emergence at $0.93R_\odot$. Case 2TwP is the most interesting since it is still influenced by convection (as we saw in the previous section) without completely loosing its characteristic bipolar structure when the loop emerges at the top of our domain. 
The results of these extra calculations are shown in Figure \ref{figure_convlat}. Both the emerging radial magnetic field and a measure of the tilt angle are represented, similar to Figure \ref{figure_shsl4}. We argue that some effects of the differential rotation are visible here. First, the rise velocities are different from one case to another, like the isentropic simulations but the differences are even more pronounced here. To be more precise, the characteristic rise time for the loop initially located at a latitude of $60^o$ is around $8$ days, when it is of more than $11$ days for the case at latitude $=15^o$. The figures represent the emerging region about $1$ to $2$ days after the first signs of emergence at $r=0.93R_\odot$. This slow down of the loops located at lower latitudes is related to the rotation rate which is stronger at lower latitudes. Indeed, as shown already in \citet{Jouve07}, the modified buoyancy force in the presence of rotation is proportional to $\Delta \rho (g-r \sin^2\theta \Omega^2)$ and the rise velocity of the loop will thus be less at low latitudes and high rotation rates, which is typically the situation for Case 2TwP15 represented here in the first panel of Figure \ref{figure_convlat}. A second effect of the differential rotation which can be noted concerns the longitude of emergence. In these differentially rotating cases, the loop which was introduced at higher latitudes has drifted in longitude by about $30^o$ (it was introduced at $100^o$ and emerges at $70^o$) when the loop in Case2TwP15 has only drifted by $5^o$. This is quite different from what we had in the isentropic cases where the drift in longitude was more important for the case at low latitudes. This was due to the stronger asymmetry between the two legs of the loop which appeared in the low latitude case. Here, the different drift values could well be due again to the differential rotation. Indeed, since the higher (lower) latitudes rotate retrograde (prograde) compared to the rotating frame, it is not surprising that the loop at high latitudes will be globally  shifted at lower longitudes compared to the low latitude cases. This is thus a clear effect here of the large-scale differential rotation acting on our rising $\Omega$-loops. Finally, the idea here was to compare the tilt angles we could get in these convective cases and compare those values to observations. It is more complicated here to identify well-defined concentrations of opposite polarities, since the convective motions have influenced the loops during their rise to sometimes produce several concentrations of radial magnetic field of both polarities. However, if we consider that the tilt angle is defined as the angle between the East-West direction and the line joining the maximum of both polarities, we are still able to measure a tilt in those 3 cases. We find a value of about $20^o$ for Case2TwP15, $40^o$ for Case2TwP and $45^o$ for Case2TwP60. The trend is thus still to increase with increasing latitude but it is less obvious than in the isentropic cases. We note here that the tilt angle distribution in the simulations of \citet{Weber11} was also much more spread in the convective cases than in the isentropic ones, which is consistent with our simulations. This is an indication here that it is not only the rotation and the associated Coriolis force acting on the loop which will determine the tilt angle observed at the solar surface but horizontal buffeting by convection could also strongly modify this measurement, even if the trend compatible with Joy's law is still found in those cases.

\section{Discussion and conclusions}
\label{sect_conclu}

In this work, we presented the dynamical evolution of twisted $\Omega$-loops embedded into a convectively stable and unstable unmagnetized environment, to mimic the behavior of a strong toroidal magnetic field generated by dynamo action at the base of the solar convection zone and rising toward the photosphere. Those simulations allow to take into account the full effects of sphericity, mean flows and convective motions on the dynamical evolution of a flux rope with a finite radius and thus contains much more physics than previous thin flux tube calculations. The first part of this work was dedicated to establish the important parameters to get a coherent radial rise in the isentropic case. We find that a loop with an initial magnetic field of $5\times 10^4 \rm G$ and with an initial twist of less than 1 turn along the loop extension ($15^o$ here) is able to rise and emerge at the top of our domain to create bipolar regions. Asymmetries develop between the leading and trailing legs of the loop due to the combined effect of angular momentum conservation and Lorentz forces and it is found that this asymmetry is strongly reduced when magnetic structures as strong as $1.5\times 10^5\rm G$ are considered. The emerging regions obtained in these isentropic cases have several properties in agreement with observations. In particular, well-identified tongues develop during emergence, mainly due to the initial twist of the magnetic field lines, also seen in observations \citep[e.g.][]{Lopez00}. For the first time, we point the appearance during flux emergence of sharp ring-shaped magnetic structures around the bipolar region, that we call ``magnetic necklaces'' and which are due to the creation of vorticity at the loop periphery. They are also very often seen in observations of large-scale flux emergence (especially in \citealt{Liu06} with MDI data) and are very similar in shape to the ones we get in our simulated active regions. Moreover, we show here that it is possible to get a case where the initial left-handed twist is strong enough to maintain the coherence of the loop and where the tilt angle of the emerging radial field agrees with Joy's law (in contrast to what is found in \citealt{Fan08}). To do so, the rise time needs to be long enough so that the Coriolis force acts sufficiently on the loop to rotate the two opposite polarities clockwise and then produce a final tilt angle of the correct sign. 

We note that if much thinner flux tubes were considered, the various parameters quoted here to get bipolar regions with properties in agreement with observations would probably have to be revised. The results do not change significantly if a tube four times smaller is introduced as long as the magnetic diffusivity is also reduced to keep a rise time small compared to the diffusion time of the magnetic structure \citep[see][for a discussion on the effect of diffusion on this type of simulations]{Jouve09}. However, if much thinner tubes are introduced (say 10 to 20 times smaller as can be expected in the real Sun), it is likely that the results may differ. In particular, the drag force coming from the pressure difference between the up and down stream sides of the turbulent wake behind the rising loop may be much bigger when the loop radius is reduced. This was shown in \citet{Batchelor67} for incompressible flows passing a rigid cylinder under large Reynolds number conditions and shown to be still true for two-dimensional simulations of buoyant loops \citep{Emonet98}. This will in turn modify the rise velocity and the competition between magnetic fields and convective motions. For example, a smaller rise velocity leaves more time for the Coriolis force to act but at the same time, an increased drag directly opposes the tilting motion, resulting in a net decrease of the tilt angle of loops of smaller radii, as shown by \citet{DSilva93}. This work would then benefit from studying the effect of loops of smaller radii but this question was not addressed here. \citet{Pinto12} also find that the wake structure is modified by the presence of a background dynamo field, possibly due to
continuous field reconnection as the tube rises.

In the convective cases, the rise velocity and characteristics of emerging regions are strongly affected by the convective motions when loops of less than $10^5 \rm G$ are considered. However, emerging regions with the correct orientation and a dominant leading polarity are still found in these simulations where the full effects of convection, rotation and sphericity are taken into account. Making direct comparison with observed photospheric field intensity in active regions is not straightforward since the scaling of the magnetic field strength with density (which is linear in our cases) may change drastically when reaching the uppermost layers of the convection zone \citep{Cheung10}. However, local helioseismology techniques seem to start to provide us with constraints on the field intensity deeper down \citep{Ilonidis11,Braun12,Ilonidis12}, much closer to the top of our computational domain. We may well soon be able to relate much more directly the results of our three-dimensional simulations to actual observations.

If the origin of the tilt angle of active regions has not been completely related to one particular property of the emerging loops yet, it is however undeniable that the initial twist of the field lines is an important ingredient. It is thus necessary to understand the generation not only of buoyant magnetic field at the base of the convection zone but of \emph{twisted} buoyant structures. It is believed that such twisted loops could be created by the advection of poloidal field lines by the initially rising strong toroidal field subject to magnetic buoyancy instabilities \citep{Longcope96, Favier12}. Full MHD simulations of dynamo-generated toroidal field becoming buoyant, gaining some twist and being able to rise coherently to the top of the convection zone are in progress \citep{Nelson11, Nelson12}. We moreover note that such simulations are able to mimic the dynamical evolution of a magnetic structure from the base of the convection zone to about $30$ Mm under the photosphere and that emergence higher up is not possible in these global anelastic simulations. Contact with three-diemnsional compressible local simulations of emerging flux in the solar atmosphere \citep{Archontis10} should now be considered to have a full picture of the MHD evolution of large-scale magnetic regions.   

When convection is considered, the emerging radial field may lose the typical characteristics of a well-defined bipolar structure that we get in the isentropic cases. Indeed, in Case 1TwP where the evolution was dominated by the downflows and upflows of the surrounding convection, the emerging structure is very complex and regions of mixed-polarity are self-consistently formed.
The large variety of large-scale emerging regions at the solar photosphere may then well be the results of magnetic $\Omega$-loops significantly processed by convection, even if robust statistical properties such as tilt angle or asymmetry between polarities indicate that these loops are still able to keep their coherence during their rise. Another possibility leading to the emergence of complex active regions in the Sun could be that such buoyant structures might interact in the bulk of the convection zone. Depending on which kind of reconnection could occur between the $\Omega$-loops, either complex region with mixed polarity and mixed helicity could emerge (such as the one observed by \citealt{Chandra10}), or the magnetic field could completely change its trajectory, leading to no emergence at all (see \citealt{Linton05} for a discussion of reconnection between twisted flux tubes). This is the analysis we intend to focus on in the near future. Further no background dynamo field was present in the simulation. \citet{Pinto12} have started to address the influence of such a dynamo field on a uniformly buoyant magnetic rope. We intend to carry on this analysis for $\Omega$-loops.

\acknowledgments
We thank the organizers of the Flux Emergence Workshops held in
St Andrews in 2007 June, in Kyoto in 2008 October, and in Berkeley in 2011 August. We also acknowledge financial support by the ERC through grant 207430 STARS2 and fundings by the Programme National Soleil Terre (INSU/PNST). We also appreciate the GENCI supercomputing centers for granting access to their infrastructures through project 1623.

\end{document}